\documentclass[aps,twocolumn,pra]{revtex4-1}
\usepackage[T1]{fontenc}
\usepackage{amsmath}
\usepackage{amssymb}
\usepackage{graphicx}
\usepackage{multirow}
\usepackage{amsfonts}
\usepackage{booktabs}
\usepackage{hyperref}
\newcommand{\abs}[1]{\lvert #1 \rvert}
\begin{document}

\title{Collapse and control of the MnAu$_{2}$ spin-spiral state
  through pressure and doping}

\author{J.~K.~Glasbrenner}
\affiliation{National Research Council/Code 6393, Naval Research
  Laboratory, Washington, DC 20375, USA}

\begin{abstract}
  MnAu$_{2}$ is a spin-spiral material with in-plane ferromagnetic Mn layers
  that form a screw-type pattern around a tetragonal $c$ axis. The spiral angle
  $\theta$ was shown using neutron diffraction experiments to decrease with
  pressure, and in later studies it was found to suffer a collapse to a
  ferromagnetic state above a critical pressure, although the two separate
  experiments did not agree on whether this phase transition is first or second
  order. To resolve this contradiction, we use density functional theory
  calculations to investigate the spiral state as a function of pressure, charge
  doping, and also electronic correlations via a Hubbard-like $U$. We fit the
  results to the one-dimensional $J_{1} - J_{2} - J_{3} - J_{4}$ Heisenberg
  model, which predicts either a first- or second-order spiral-to-ferromagnetic
  phase transition for different regions of parameter space. At ambient
  pressure, MnAu$_{2}$ sits close in parameter space to a dividing line
  separating first- and second-order transitions, and a combination of pressure
  and electron doping shifts the system from the first-order region into the
  second-order region. Our findings demonstrate that the contradiction in
  pressure experiments regarding the kind of phase transition are likely due to
  variations in sample quality. Our results also suggest that MnAu$_{2}$ is
  amenable to engineering via chemical doping and to controlling $\theta$ using
  pressure and gate voltages, which holds potential for integration in
  spintronic devices.
\end{abstract}

\maketitle

\section{Introduction}
\label{sec:introduction}

The tetragonal material MnAu$_{2}$ is one of the oldest known spin-spiral
materials \cite{Meyer1956_JPhysRadium_Proprietes,
  Herpin1959_ComptesRendus_Structure, Herpin1961_JPhysRadium_Etude}, with a
N\'{e}el temperature of $T_{N} = 363 \text{ K}$
\cite{Meyer1956_JPhysRadium_Proprietes} and a local Mn moment of $3.5 \mu_{B}$
\cite{Herpin1961_JPhysRadium_Etude, Samata1998_JPhysChemSol_Giant}. The magnetic
ground state consists of in-plane ferromagnetic Mn layers with a screw-type
pattern around the tetragonal $c$ axis, which can be described using the spiral
angle $\theta$ [this is equivalent to the wave vector
$\textbf{q} = (0, 0, q_z)$, where $\theta = q_{z}c/2$]. Neutron diffraction
measurements \cite{Herpin1961_JPhysRadium_Etude, Smith1966_JPhysChemSol_neutron,
  Handstein2005_JMag} find that $\theta$ has a weak dependence on temperature,
increasing slightly with increasing temperature
[$\theta (77 \text{ K}) = 47^{\circ}$ and
$\theta (295 \text{ K}) = 51^{\circ}$]. The spiral state is known to collapse to
a metamagnetic fan-like configuration when placed in a $\sim 10 \text{ kOe}$
magnetic field at room temperature \cite{Meyer1956_JPhysRadium_Proprietes,
  Herpin1959_ComptesRendus_Structure}, which led to a revival of interest in
MnAu$_{2}$ when it was found that this gives rise to a giant magnetoresistive 
effect \cite{Samata1998_JPhysChemSol_Giant}.

There are several microscopic mechanisms that can induce spin spirals. In
itinerant systems, both Fermi surface nesting
\cite{Evenson1968_PhysRevLett_Generalized} and the Ruderman-Kittel-Kasuya-Yosida
(RKKY) interaction \cite{Ruderman1954_PhysRev_Indirect,
  Kasuya1956_ProgTheorPhys_Theory, Yosida1957_PhysRev_Magnetic} can lead to
spiral formation. The Dzyaloshinsky-Moriya (DM) interaction
\cite{Dzyaloshinsky1958_JPhysChemSol_thermodynamic,
  Moriya1960_PhysRev_Anisotropic}, which occurs in materials without an
inversion center (MnAu$_{2}$ has an inversion center, which rules it out as a
possible mechanism), can also induce spirals and moment canting, but due to the
interaction's relativistic origin it is more important in materials with heavy
elements. It can still be competitive in lighter transition metals, such as MnSi
\cite{Bak1980_JPhysC_Theory, Nakanishi1980_SolStComm_origin,
  Kataoka1981_JPhysSocJpn_Helical}, but the spirals will have a long wavelength
due to weak relativistic effects. Finally, frustration, which can be geometric
or magnetic, can encourage noncollinearity and lead to the formation of spirals.

The spin spirals in MnAu$_{2}$ are understood to be a consequence of magnetic
frustration, which is commonly modeled using a one-dimensional Heisenberg model
with first and second neighbor interplanar couplings (the $J_{1}-J_{2}$ model)
\cite{Enz1961_JApplPhys_Magnetization, Nagamiya1962_JApplPhys_Modification}. In
this model, the spiral state is stable when $J_{1} \neq 0$, $J_{2} > 0$, and
$\abs{J_{1}} < 4 \abs{J_{2}}$ \cite{Enz1961_JApplPhys_Magnetization}. The
validity of applying this model to MnAu$_{2}$ was confirmed using density
functional theory (DFT) calculations \cite{Udvardi2006_PhysRevB_Helimagnetism,
  Glasbrenner2014_PhysRevB_Magnetic}. The origin of the magnetic frustration was
traced to a competition between two exchange mechanisms, nearest-neighbor
superexchange and a transferred RKKY-like interaction, and electronic
correlations were found to play an essential role in suppressing the RKKY-like
interaction, which is necessary to satisfy the $\abs{J_{1}} < 4 \abs{J_{2}}$
inequality \cite{Glasbrenner2014_PhysRevB_Magnetic}.

The spiral angle $\theta$ is sensitive to pressure, with neutron diffraction
measurements determining that $\theta (P)$ decreases with applied pressure [
$\theta (0) = 50.7^{\circ}$ ($47.0^{\circ}$) decreasing to
$\theta (8.83 \text{ kbar}) = 41.8^{\circ}$ ($40.5^{\circ}$) at a temperature of
295 K (77 K)] \cite{Smith1966_JPhysChemSol_neutron}. Measurements of the
critical magnetic field $H_{c}$ for the spiral-to-ferromagnetic transition as a
function of pressure, when extrapolated to zero external field, implied that the
spiral state should collapse to ferromagnetism at $\sim 12$ kbar
\cite{Grazhdankina1963_SovPhysJETP_Effect}. The pressure-induced
spiral-to-ferromagnetic transition was confirmed by subsequent experiments using
inductance measurements \cite{Wayne1969_JPhysChemSol_pressure}, electrical
resistivity measurements, and measurements of $H_{c}$
\cite{Adiatullin1971_SovPhysSolSt_Influence}, although the two reports disagreed
on the order of the phase transition. In
Ref.~\onlinecite{Wayne1969_JPhysChemSol_pressure}, the transition is of the
second kind and occurs over a pressure range of 12-20 kbar, while in
Ref.~\onlinecite{Adiatullin1971_SovPhysSolSt_Influence} it is of the first kind
and occurs at 13 kbar. This contradiction was never resolved and remained an
open question.

The $J_{1}-J_{2}$ Heisenberg model, which can explain the stability of the
spiral state, predicts that the transition must be of the second kind. However,
general investigations of the classical one-dimensional Heisenberg model show
that including the third nearest-neighbor term $J_{3}$ leads to regions of
parameter space with first order phase transitions
\cite{Pimpinelli1989_JPhysCondensMatter_Classical,
  Zinke2010_JPhysCondensMatter_Spiral}. Further neighbors, such as $J_{4}$,
change the area of parameter space where first-order transitions occur
\cite{Zinke2010_JPhysCondensMatter_Spiral}. It follows then that if a system
were to sit near the boundary in parameter space separating first- and
second-order phase transitions, then variations in material quality could shift
the system towards one region or the other.

We argue that this is the case for MnAu$_{2}$. We support this claim using DFT
calculations to simulate the effects of pressure, charge doping, and variation
in the strength of electronic correlations via a Hubbard-like $U$. We calculate
the magnetic energies and fit them with a classical, one-dimensional Heisenberg
model with coupling between the first four nearest-neighbor Mn planes. We
confirm that ideal, stoichiometric MnAu$_{2}$ sits close to a boundary
separating first- and second-order phase transitions, and that varying the
pressure and doping shifts MnAu$_{2}$ across this critical boundary. This
suggests that the two pressure experiments are not in conflict, but instead the
MnAu$_{2}$ samples from each are located in two separate portions of phase
space. The spiral angle $\theta$ is also found to be responsive to applied
pressure and doping, which holds promise for integration of MnAu$_{2}$ into
spintronic devices.

\begin{figure*}[t]
  \centering
  \includegraphics[width=0.85\textwidth]{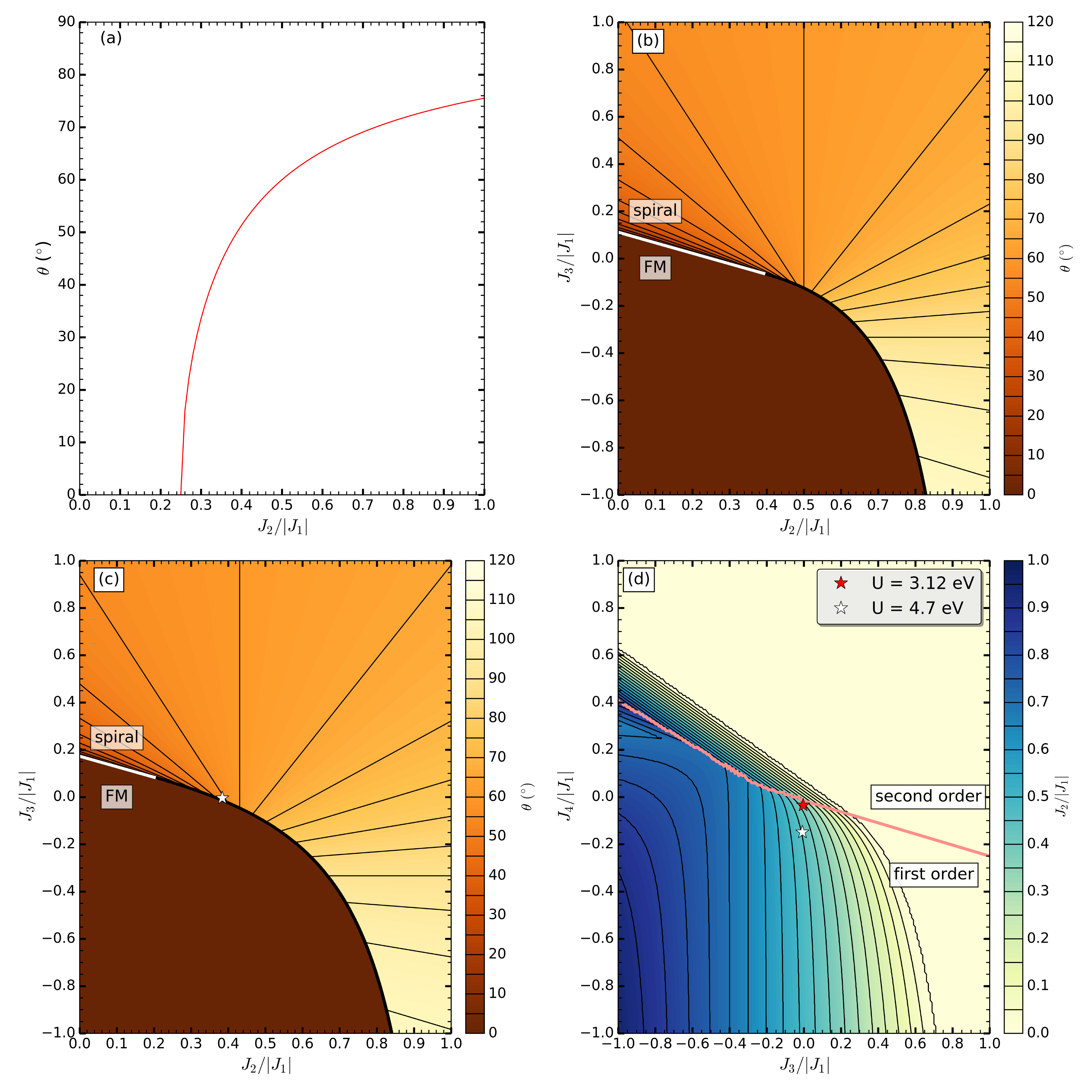}
  \caption{(Color online) Phase diagrams for the one-dimensional Heisenberg
    model of Eq.~\eqref{eq:1}. (a) The phase diagram of the $J_{1} - J_{2}$
    model. (b) The phase diagram of the $J_{1} - J_{2} - J_{3}$ model, where the
    contour plot corresponds to $\theta$. The thick black line indicates a
    first-order phase transition and the thick white line indicates a
    second-order transition. (c) The phase diagram of the
    $J_{1} - J_{2} - J_{3} - J_{4}$ model for $J_{4} / \abs{J_{1}} = -0.0346$,
    plotted in the same manner as panel (b).  The white star indicates where
    ideal stoichiometric MnAu$_{2}$ with $U = 3.12 \text{ eV}$ sits in the phase
    diagram. (d) The critical values of $J_{2} / \abs{J_{1}}$ (contour plot),
    $J_{3} / \abs{J_{1}}$, and $J_{4} / \abs{J_{1}}$ for the
    spiral-to-ferromagnetic phase transition. The red line divides parameter
    space into two regions, where first-order phase transitions occur north of
    the line, and second-order transitions occur south of it. The red and white
    stars represent where ideal, stoichiometric MnAu$_{2}$ sits for
    $U = 3.12 \text{ eV}$ and $U = 4.7 \text{ eV}$, respectively.}
  \label{fig:heisenberg_model}
\end{figure*}

\section{Computational Details}
\label{sec:comp-deta}

We employed DFT to solve the electronic structure of MnAu$_{2}$ and calculate
the total energy of magnetic configurations. To perform these calculations, we
used the all-electron code \textsc{elk} \cite{elk}, which is an implementation
of the full-potential linear augmented planewave method, and also projector
augmented wave potentials implemented in the code \textsc{vasp}
\cite{Kresse1993_PhysRevB_initio, Kresse1996_PhysRevB_Efficient}. We used the
local spin-density approximation (LSDA) \cite{Perdew1992_PhysRevB_Accurate} for
our calculations, and for correlation effects we used the DFT$+U$ method in the
fully localized limit \cite{Liechtenstein1995_PhysRevB_Densityfunctional}, in
which we introduce a Hubbard-like $U$ on the $3d$ orbitals of the Mn atoms. We
used two values of $U$, $U = 3.12 \text{ eV}$ and $4.7 \text{ eV}$, and $J$ was
set to $0.7 \text{ eV}$.

MnAu$_{2}$ has a tetragonal crystal structure with $I4/mmm$ space group
symmetry. The experimental lattice parameters are $a = 3.370$ \AA~and
$c/a = 2.599$, yielding a volume of 49.741 \AA$^3$/formula unit, and the Wyckoff
positions for Mn and Au are $2a$ and $4e$, respectively, with the internal
experimental height for Au $z_{Au} = 0.34$ (fractional coordinates)
\cite{Adachi1972_JPhysSocJpn_Helical, Hart1970_JApplPhys_Magnetic}. To simulate
pressure, we began with the MnAu$_{2}$ unit cell from experiment with tetragonal
symmetry and varied the lattice parameters $a$ and $c$ to set the volume
\footnote{The experimental pressure studies of MnAu$_{2}$ do not find changes in
  the crystal structure prior to the spiral-to-ferromagnetic phase
  transition. For this reason, we only investigate tetragonal unit cells and do
  not consider whether other structures become competitive at higher pressures,
  since they will not impact the phase transition.}. We then fixed the volume
and performed structural relaxations in \textsc{vasp} to optimize the $c/a$
ratio and the internal parameter $z_{Au}$. After the relaxation, we performed
total energy calculations in \textsc{vasp} with the tetrahedron method and
calculated the pressure using the formula $P = - dE/dV$. We then imported the
relaxed structures into \textsc{elk} for calculating the magnetic energies.

We employed two methods to simulate charge doping, (1) adding electrons directly
to the system along with a uniform positive charge background (hereafter
referred to as ``charge dosing''), and (2) using the virtual crystal
approximation (VCA). The VCA involves replacing the Au ions with fictitious ions
of fractional charge in order to add electrons or holes to the system. We used
the experimental lattice parameters of MnAu$_{2}$ in our doping calculations,
while for calculations with simultaneous charge doping and applied pressure we
used the relaxed cells obtained with \textsc{vasp}.

We used the spin spiral method implemented in \textsc{elk} to calculate the
total energy as a function of the spiral angle. This method defines a wave
vector $\mathbf{q} = (0, 0, q_z)$ that is applied to a primitive cell to
simulate spirals. In our previous study of MnAu$_{2}$
\cite{Glasbrenner2014_PhysRevB_Magnetic}, we found good agreement between the
calculated energies obtained via this method and noncollinear calculations of
spirals in supercells. In this study, our procedure was to calculate the energy
as a function of the wave vector for the range $0 < q_{z} < 2 \pi/c$ as we
varied the pressure, charge doping, and Hubbard $U$. We then used
$\theta = q_{z} c / 2$ to convert the wave vector to an angle.

Convergence for the spin spiral calculations was carefully checked in the
primitive cells, and the following parameters were necessary to achieve
convergence: k-mesh = $16 \times 16 \times 16$, nempty = 8, rgkmax = 8.0, gmaxvr
= 12, lmaxapw = 8, and the smearing function width was set to 0.0001 Ha.

\section{Summary of 1D Heisenberg Model}
\label{sec:summ-1d-heis}

\begin{figure}[t]
  \centering
  \includegraphics[width=0.48\textwidth]{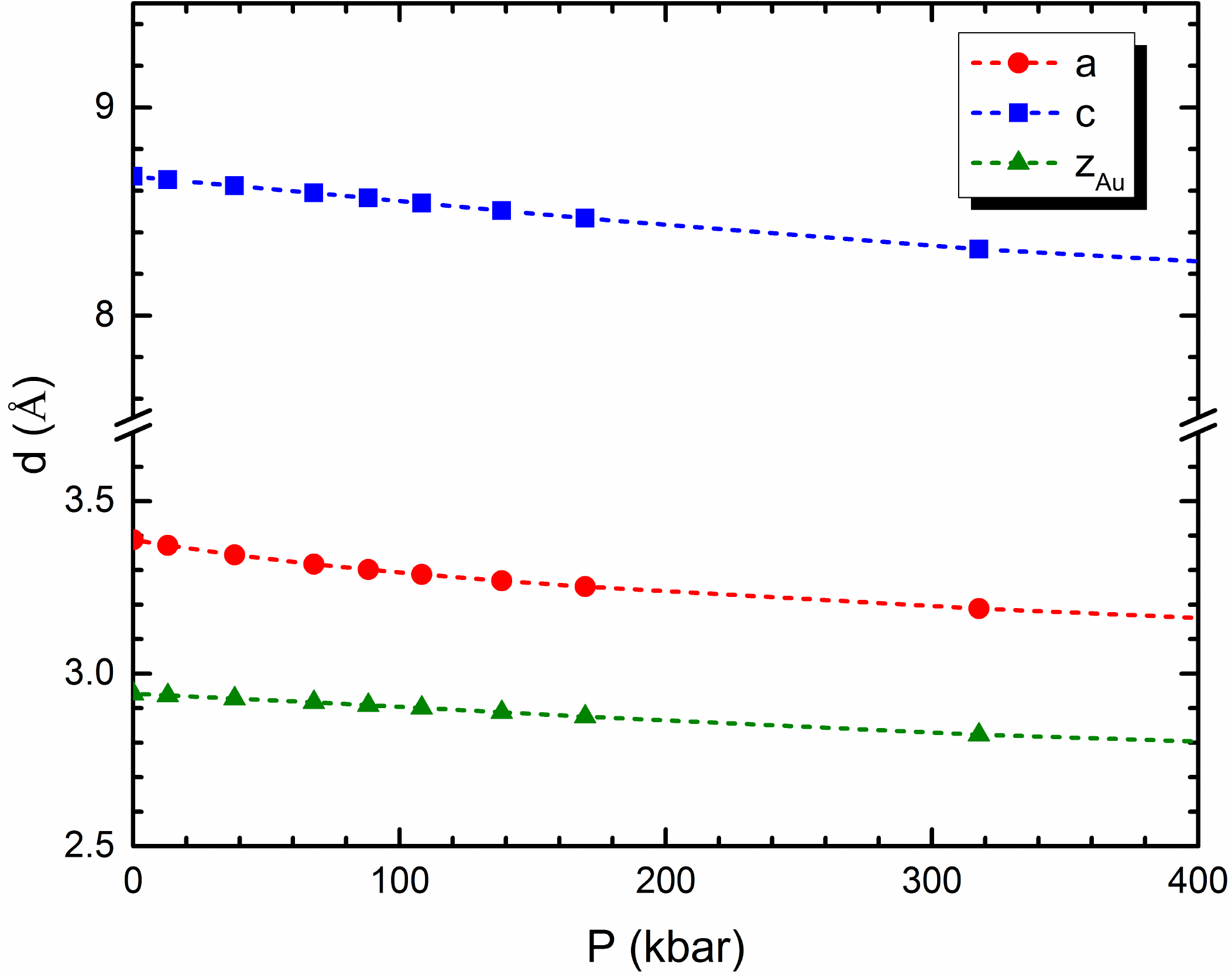}
  \caption{(Color online) MnAu$_{2}$ lattice parameters as a function of
    pressure.}
  \label{fig:lattice}
\end{figure}

\begin{figure*}[t]
  \centering
  \includegraphics[width=0.95\textwidth]{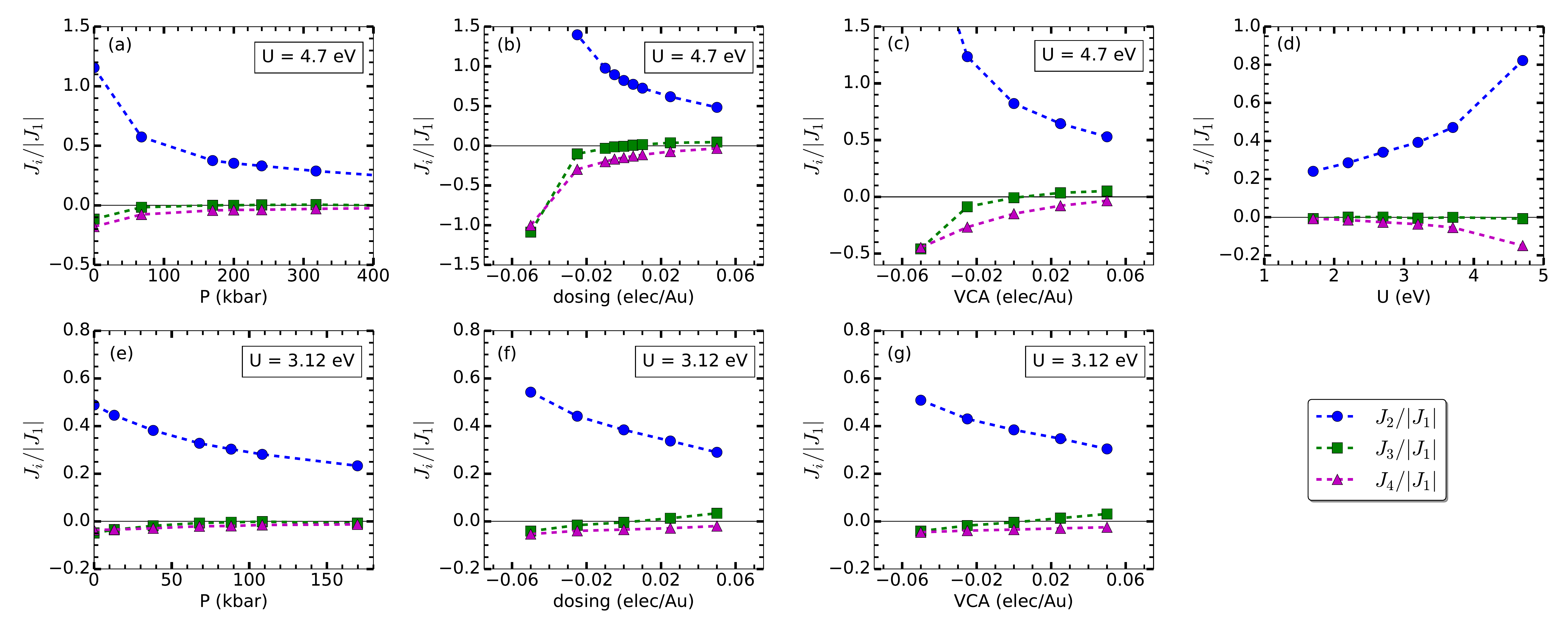}
  \caption{(Color online) The dependence of the exchange parameters
    $J_{2}/\abs{J_{1}}$, $J_{3}/\abs{J_{1}}$, and $J_{4}/\abs{J_{1}}$, see
    legend, as a function of pressure, doping, and $U$. In the charge dosing and
    VCA panels, positive values on the horizontal axis correspond to electron
    doping and negative values to hole doping. (a) Pressure dependence with
    $U = 4.7 \text{ eV}$. (b) Charge dosing dependence with
    $U = 4.7 \text{ eV}$. (c) VCA dependence with $U = 4.7 \text{ eV}$. (d)
    Dependence on Hubbard $U$. (e) Pressure dependence with
    $U = 3.12 \text{ eV}$. (f) Charge dosing dependence with
    $U = 3.12 \text{ eV}$. (g) VCA dependence with $U = 3.12 \text{ eV}$.}
  \label{fig:exchangeparams}
\end{figure*}

The magnetic interactions in MnAu$_{2}$ are typically modeled using the
Heisenberg model:
\begin{align}
  \label{eq:1}
  H &= \sum_{i \neq j} J_{ij} \hat{\mathbf{m}}_{i} \cdot \hat{\mathbf{m}}_{j}
\end{align}
Note that $J_{ij} = \bar{J}_{ij} \abs{\mathbf{m}}^2$, where $\bar{J}_{ij}$ is
the magnetic coupling and $\abs{\mathbf{m}}$ is the moment amplitude. DFT
calculations confirm that Mn is in the high spin state with $S = 5/2$, which
makes the classical Heisenberg model suitable for studying the magnetic
interaction. The magnetic moment is also assumed to be constant, which is
reasonable as the Mn moment amplitudes do not vary by much as a function of
$\theta$.

The spiral state of MnAu$_{2}$ is completely described by the parameter
$\theta$. As a result, we can simplify the Heisenberg model to one dimension:
\begin{align}
  \label{eq:2}
  H &= \text{const.} + \sum_{n = 1}^{N} J_{n} \cos(n \theta)
\end{align}
The interplanar coupling determines the spiral state stability, and below we
discuss the phase diagram for the models which include two ($J_{1} - J_{2}$
model), three ($J_{1} - J_{2} - J_{3}$ model), and four
($J_{1} - J_{2} - J_{3} - J_{4}$ model) nearest-neighbors in Eq.~\eqref{eq:2}.

The $J_{1} - J_{2}$ model is the simplest possible model with a stable spiral
state, which occurs when $J_{1} \neq 0$, $J_{2} > 0$, and
$\abs{J_{1}} < 4 \abs{J_{2}}$. The angle $\theta$ is plotted as a function of
$J_{2} / \abs{J_{1}}$ in Fig.~\ref{fig:heisenberg_model}(a), and the
spiral-to-ferromagnetic phase transition is of second order. What has not been
appreciated in previous studies of MnAu$_{2}$ is that this model cannot explain
how the turn angle could collapse under pressure in a first-order transition, as
measured in Ref.~\onlinecite{Adiatullin1971_SovPhysSolSt_Influence}.

Including the third nearest-neighbor coupling yields the $J_{1} - J_{2} - J_{3}$
model, which can be solved analytically (see Supplemental Material
\cite{supp_mater}) \cite{Pimpinelli1989_JPhysCondensMatter_Classical,
  Zinke2010_JPhysCondensMatter_Spiral}. The phase diagram of $\theta$ for this
model as a function of $J_{2} / \abs{J_{1}}$ and $J_{3} / \abs{J_{1}}$ is
depicted in Fig.~\ref{fig:heisenberg_model}(b). There are two regions in the
parameter space, the ferromagnetic region in the lower left and the spiral
region in the rest of the diagram, with the contour plot corresponding to the
magnitude of $\theta$. The phase boundaries separating the two regions are
\cite{Pimpinelli1989_JPhysCondensMatter_Classical}:
\begin{align}
  \label{eq:j1j2j3boundary}
  J_{3} &= \begin{cases}
    \frac{1}{9} \left(\abs{J_{1}} - 4 J_{2} \right) & 0 \leq J_{2} / \abs{J_{1}} \leq 2 / 5 \\
    \frac{J_{2}^2}{4 \left(J_{2} - \abs{J_{1}} \right)} & J_{2} / \abs{J_{1}} > 2 / 5
    \end{cases}.
\end{align}
Importantly, this model contains both first- and second-order phase transitions,
which are distinguished by the kind of degeneracy that occurs on the borders in
Eq.~\eqref{eq:j1j2j3boundary}. A first-order phase transition occurs when there
is a degeneracy between the ferromagnetic solution ($\theta = 0$) and a finite
angle ($\theta > 0$), while the second order phase transition corresponds to a
smooth, continuous connection between ferromagnetic and spin-spiral regions, and
is found by taking a second-order Maclaurin series expansion of Eq.~\eqref{eq:2}
and equating it with the ferromagnetic solution in the limit $\theta \to
0$. Second-order transitions correspond with the
$0 \leq J_{2} / \abs{J_{1}} \leq 2 / 5$ result and are depicted as the thick
white line in Fig.~\ref{fig:heisenberg_model}(b), while first order transitions
correspond to the $J_{2} / \abs{J_{1}} > 2 / 5$ result and are depicted as the
thick black line.  From this, it is clear that a small ferromagnetic coupling
between third-neighbor planes (relative to $\abs{J_{1}}$) can change the order
of the phase transition.

The analysis is more complicated for four or more neighbor couplings since an
analytic solution for $\theta$ is not available. Despite this, some properties
can still be analytically solved (see Supplemental Material \cite{supp_mater})
\cite{Zinke2010_JPhysCondensMatter_Spiral}. The phase boundary for second order
transitions is $J_{2} = \left(\abs{J_{1}} - 9 J_{3} - 16 J_{4} \right) / 4$. The
criterion for a first order phase transition is
\begin{align}
  \label{eq:4}
  J_{3} / \abs{J_{1}} < - \frac{1 + 64 J_{4} / \abs{J_{1}}}{15}.
\end{align}
Note that the expression for the second order phase boundary and
Eq.~\eqref{eq:4} are valid when $J_{3}, J_{4} > 0$. If one of the parameters is
negative, then the expressions still hold as long as the antiferromagnetic
couplings dominate \cite{Zinke2010_JPhysCondensMatter_Spiral}, such as
$J_{3} > 0$, $J_{4} < 0$, and $\abs{J_{3}} \gg \abs{J_{4}}$. What's clear is
that a relatively small fourth neighbor coupling can have a dramatic effect on
the location of the first order and second order phase boundaries. To illustrate
this, we numerically calculate the phase diagram for the
$J_{1} - J_{2} - J_{3} - J_{4}$ model with $J_{4} / \abs{J_{1}} = -0.0346$,
which we plot in Fig.~\ref{fig:heisenberg_model}(c). Comparing panels (b) and
(c), we clearly see that the length of the second order phase boundary (white
line) has changed, despite the relative weakness of the fourth neighbor
coupling. We also calculate the exchange parameters for ideal, stoichiometric
MnAu$_{2}$ with $U = 3.12 \text{ eV}$, which are $J_{2} / \abs{J_{1}} = 0.3840$,
$J_{3} / \abs{J_{1}} = -0.0043$, $J_{4} / \abs{J_{1}} = -0.0346$, and indicate
where MnAu$_{2}$ sits with the white circle. The material sits very close to the
first order transition, and a small variation in $J_{3} / \abs{J_{1}}$ would
drive the system towards ferromagnetism.

In Fig.~\ref{fig:heisenberg_model}(d) we plot a contour map of the critical
values of $J_{2} / \abs{J_{1}}$, $J_{3} / \abs{J_{1}}$, and
$J_{4} / \abs{J_{1}}$ for the spiral-to-ferromagnetic transition in the
one-dimensional $J_{1} - J_{2} - J_{3} - J_{4}$ model. We only consider
$J_{2} \geq 0$ in this figure. The contour plot shows the surface in
$(J_{2} / \abs{J_{1}}$, $J_{3} / \abs{J_{1}}$, $J_{4} / \abs{J_{1}})$ space that
separates ferromagnetic and spiral order. For example, if you calculate the
exchange parameters for MnAu$_{2}$ and those parameters place the system
``inside'' the surface $(J_{2} < J_{2}^{c})$, then the ground state magnetic
order is ferromagnetic. If instead the parameters place MnAu$_{2}$ on or outside
the surface $(J_{2} \geq J_{2}^{c})$, then the magnetic order will be a spiral
state. Note that the light yellow region corresponding to $J_{2}^{c} = 0$ means
that the system has spiral order for all $J_{2} \geq 0$. Finally, if you can
tune the parameters, such as by applying pressure or adding holes/electrons,
then the system can cross the surface and undergo a phase transition.

As we saw in panels (b) and (c) of Fig.~\ref{fig:heisenberg_model}, phase
transitions can either be first or second order. The parameters
$(J_{3} / \abs{J_{1}}$, $J_{4} / \abs{J_{1}})$ control the kind of phase
transition that occurs when you cross the surface depicted in
Fig.~\ref{fig:heisenberg_model}(d). The red line separates the first- and
second-order transition regions. If you cross the surface while north of the
line, then the phase transition is second order; if you cross the surface while
south of the line, then the phase transition is first order. Having established
this, it is now instructive to indicate where ideal, stoichiometric MnAu$_{2}$
sits in $(J_{3} / \abs{J_{1}}$, $J_{4} / \abs{J_{1}})$ space for
$U = 3.12 \text{ eV}$ (red star) and $U = 4.7 \text{ eV}$ (white star). For both
systems, $J_{2} > J_{2}^{c}$, so they have spiral order. What is striking here
is how close MnAu$_{2}$ sits to the dividing line, especially when
$U = 3.12 \text{ eV}$. This shows that changes in the exchange parameters
induced through pressure, doping, or impurities could place the system north of
the red line. Since the system sits close to the dividing line, this could
explain the discrepancy in experimental pressure studies discussed in
Sec.~\ref{sec:introduction}.

One final point to note is that this analysis assumed that the local moments
always have an in-plane orientation. Neutron diffraction experiments confirm
that spin spirals in MnAu$_{2}$ are oriented in-plane
\cite{Herpin1959_ComptesRendus_Structure}, and our calculations of the
magnetocrystalline anisotropy energy for MnAu$_{2}$ with $U = 3.12$ eV give
$E(001) - E(100) = 0.35 \text{ meV}$ \footnote{Convergence of the
  magnetocrystalline anisotropy energy in the MnAu$_{2}$ unit cell was achieved
  with the following parameters in \textsc{elk}: k-mesh =
  $28 \times 28 \times 28$, nempty = 12, rgkmax = 8.0, gmaxvr = 16, lmaxapw =
  14, lmaxvr = 14, and the smearing function width was set to 0.0001 Ha},
indicating the in-plane direction is the preferred axis. Yet, it is less clear
what would happen on the first order transition boundary, where zero and finite
angle configurations are degenerate. We checked this possibility by considering
conical solutions. We added a canted angle $\varphi$ and a uniaxial
magnetocrystalline anisotropy term to Eq.~\eqref{eq:2},
\begin{multline}
  \label{eq:3}
  E(\theta, \varphi) = \text{const} + D \sin^2(\varphi) + \\ 
  \sum_{n=1}^{m} J_{n} \left[ 1 - 2 \sin^2(\varphi) \sin^2 \left( \frac{n\theta}{2} \right) \right].
\end{multline}
In this expression, the angles $\theta$ and $\varphi$ remain decoupled and the
solutions for $\theta$ are unaffected by canting and simple uniaxial
magnetocrystalline anisotropy. If we now restrict ourselves to the first-order
phase transition boundary, the energy difference between the conical spirals and
the ferromagnetic configuration is
\begin{align}
  \label{eq:5}
  E(\theta, \varphi) - E(0, \varphi') = D \left[ \sin^2(\varphi) - \sin^2(\varphi') \right]
\end{align}
Therefore in the absence of magnetocrystalline anisotropy there is a continuous
degeneracy of the canting angle. This degeneracy is lifted with the
magnetocrystalline anisotropy term, although we should note that the
magnetocrystalline anisotropy parameter is small,
$D = 0.35 \text{ meV} \sim 4 \text{ K}$, and since the pressure experiments
\cite{Smith1966_JPhysChemSol_neutron, Wayne1969_JPhysChemSol_pressure,
  Adiatullin1971_SovPhysSolSt_Influence} were performed at temperatures of 77 K
and at room temperature, it is possible for the ferromagnetic orientation to
align along the $c$ axis. It could also be used as a continuous pathway to make
$\theta$ drop to zero (at finite $\theta$ the moments can align parallel to $c$
while on the first-order boundary, and then as you move away from the boundary
they can drop back to being in-plane with $\theta = 0$). This means that the
phase transition may not be truly first order. However, it should be noted that
such a transition entails a rapid change in the canting angle, which would be
difficult to distinguish from a first-order transition in experiment. For the
rest of the discussion, we will continue to refer to this as a first-order
transition with the caveat that it may correspond to a rapid change in the
canting angle. Regardless, it remains distinct from the second-order transition,
where $\theta$ continuously decreases to zero.

\section{DFT Results and Discussion}
\label{sec:dft-results}

\begin{table*}[t]
  \centering
  \begin{tabular}[c]{c c c c c c c c c c c}
    \toprule
    Pressure & $J_{1}$ & $J_{2}$ & $J_{3}$ & $J_{4}$ & Doping & Type & $J_{1}$ & $J_{2}$ & $J_{3}$ & $J_{4}$ \\
    (kbar) & \multicolumn{4}{c}{(meV)} & (elec/Au) & & \multicolumn{4}{c}{(meV)} \\ \midrule
    \multicolumn{11}{c}{$U = 3.12 \text{ eV}$} \\ \midrule
      0.000 & -20.1394 &  9.8339 & -0.9823 & -0.7333 & -0.050 & Dosing & -12.5109 & 6.7832 & -0.5134 & -0.6678 \\
     12.977 & -22.3635 &  9.9571 & -0.7869 & -0.7692 &        & VCA    & -14.1347 & 7.1857 & -0.5786 & -0.6446 \\
     38.109 & -26.5047 & 10.1130 & -0.4949 & -0.7711 & -0.025 & Dosing & -16.2810 & 7.1840 & -0.2603 & -0.6489 \\
     67.892 & -31.1437 & 10.2032 & -0.2159 & -0.6435 &        & VCA    & -17.1181 & 7.3614 & -0.3046 & -0.6662 \\
     88.298 & -34.1544 & 10.3499 & -0.1301 & -0.6544 &  0.000 & Dosing & -20.2505 & 7.7762 & -0.0861 & -0.7000 \\
    108.353 & -37.0067 & 10.4056 & -0.0440 & -0.5633 &        & VCA    & -20.2505 & 7.7762 & -0.0861 & -0.7000 \\
    169.709 & -44.9705 & 10.4817 & -0.3415 & -0.5701 &  0.025 & Dosing & -24.5855 & 8.2909 &  0.3200 & -0.7065 \\
            &          &         &         &         &        & VCA    & -23.5647 & 8.1775 &  0.3180 & -0.6913 \\
            &          &         &         &         &  0.050 & Dosing & -29.1444 & 8.4463 &  0.9843 & -0.5759 \\
            &          &         &         &         &        & VCA    & -26.9893 & 8.2059 &  0.8370 & -0.6678 \\ \midrule
    \multicolumn{11}{c}{$U = 4.7 \text{ eV}$} \\ \midrule
      0.000 &  -5.5435 &  6.4037 & -0.6415 & -0.9926 & -0.050 & Dosing &  -0.6574 & 3.6749 & -0.7146 & -0.6583 \\
     67.892 & -13.4226 &  7.7049 & -0.2128 & -1.0301 &        & VCA    &  -1.6202 & 4.0940 & -0.7442 & -0.7241 \\ 
    169.709 & -24.2692 &  9.1221 &  0.0078 & -1.0323 & -0.025 & Dosing &  -3.1144 & 4.3517 & -0.3170 & -0.9235 \\
    200.000 & -27.4309 &  9.6846 &  0.0290 & -1.0575 &        & VCA    &  -3.6679 & 4.5317 & -0.3207 & -0.9799 \\
    240.000 & -31.4087 & 10.4061 &  0.1081 & -1.1649 & -0.010 & Dosing &  -4.8464 & 4.7346 & -0.1601 & -0.9704 \\
    317.691 & -39.3228 & 11.3247 &  0.2179 & -1.1539 & -0.005 & Dosing &  -5.4763 & 4.9005 & -0.0754 & -0.9132 \\
            &          &         &         &         &  0.000 & Dosing &  -6.1229 & 5.0350 & -0.0480 & -0.9103 \\
            &          &         &         &         &        & VCA    &  -6.1229 & 5.0350 & -0.0480 & -0.9103 \\
            &          &         &         &         &  0.005 & Dosing &  -6.7465 & 5.2203 &  0.0486 & -0.8760 \\
            &          &         &         &         &  0.010 & Dosing &  -7.4455 & 5.4008 &  0.1156 & -0.8414 \\
            &          &         &         &         &  0.025 & Dosing &  -9.6007 & 5.9304 &  0.3580 & -0.7010 \\
            &          &         &         &         &        & VCA    &  -8.8932 & 5.7378 &  0.3137 & -0.6957 \\
            &          &         &         &         &  0.050 & Dosing & -13.4470 & 6.5046 &  0.6154 & -0.4418 \\
            &          &         &         &         &        & VCA    & -11.7329 & 6.2087 &  0.6055 & -0.4044 \\ \midrule
    $U$ & $J_{1}$ & $J_{2}$ & $J_{3}$ & $J_{4}$   & & & & & & \\ 
    (eV) & \multicolumn{4}{c}{(meV)}            & & & & & & \\ \midrule
    1.7 & -37.7314 & 9.0945 & -0.2716 & -0.2806 & & & & & & \\
    2.2 & -31.4231 & 8.9775 &  0.0325 & -0.4564 & & & & & & \\
    2.7 & -25.0395 & 8.5043 &  0.0267 & -0.6542 & & & & & & \\
    3.2 & -19.4161 & 7.6273 & -0.0997 & -0.7022 & & & & & & \\
    3.7 & -14.3696 & 6.7687 & -0.0028 & -0.7722 & & & & & & \\
    4.7 &  -6.1229 & 5.0350 & -0.0480 & -0.9103 & & & & & & \\
    \bottomrule
  \end{tabular}
  \caption{Values of the exchange parameters as a function of pressure,
    charge dosing, the VCA, and $U$. Pressure and doping results for
    $U = 3.12 \text{ eV}$ and $U = 4.7 \text{ eV}$ are reported in the first
    two sections of the table, where the first five columns are the pressure
    results and the last six columns are the doping results. The ``type''
    column indicates whether the doping was performed using charge dosing or
    the VCA. The parameters as a function of $U$ are reported in the last
    section of the table.}
  \label{tab:exchange_parameters}
\end{table*}

Having established in Sec.~\ref{sec:summ-1d-heis} that MnAu$_{2}$ sits close to
a dividing line that separates first and second order phase transitions, we now
use DFT to perform a systematic set of spin-spiral calculations to simulate the
effects of pressure and doping, and the effect of varying $U$. This will reveal
how applying pressure and doping makes MnAu$_{2}$ evolve in parameter
space. Each set of spin spiral calculations results in a $E(\theta)$ curve, and
all curves are similar to what was obtained in our previous report
\cite{Glasbrenner2014_PhysRevB_Magnetic}. Each curve was fitted using the
$J_{1}-J_{2}-J_{3}-J_{4}$ model, and from these we extract the exchange
constants (see Supplemental Material \cite{supp_mater}).

\begin{figure*}[t]
  \centering
  \includegraphics[width=0.95\textwidth]{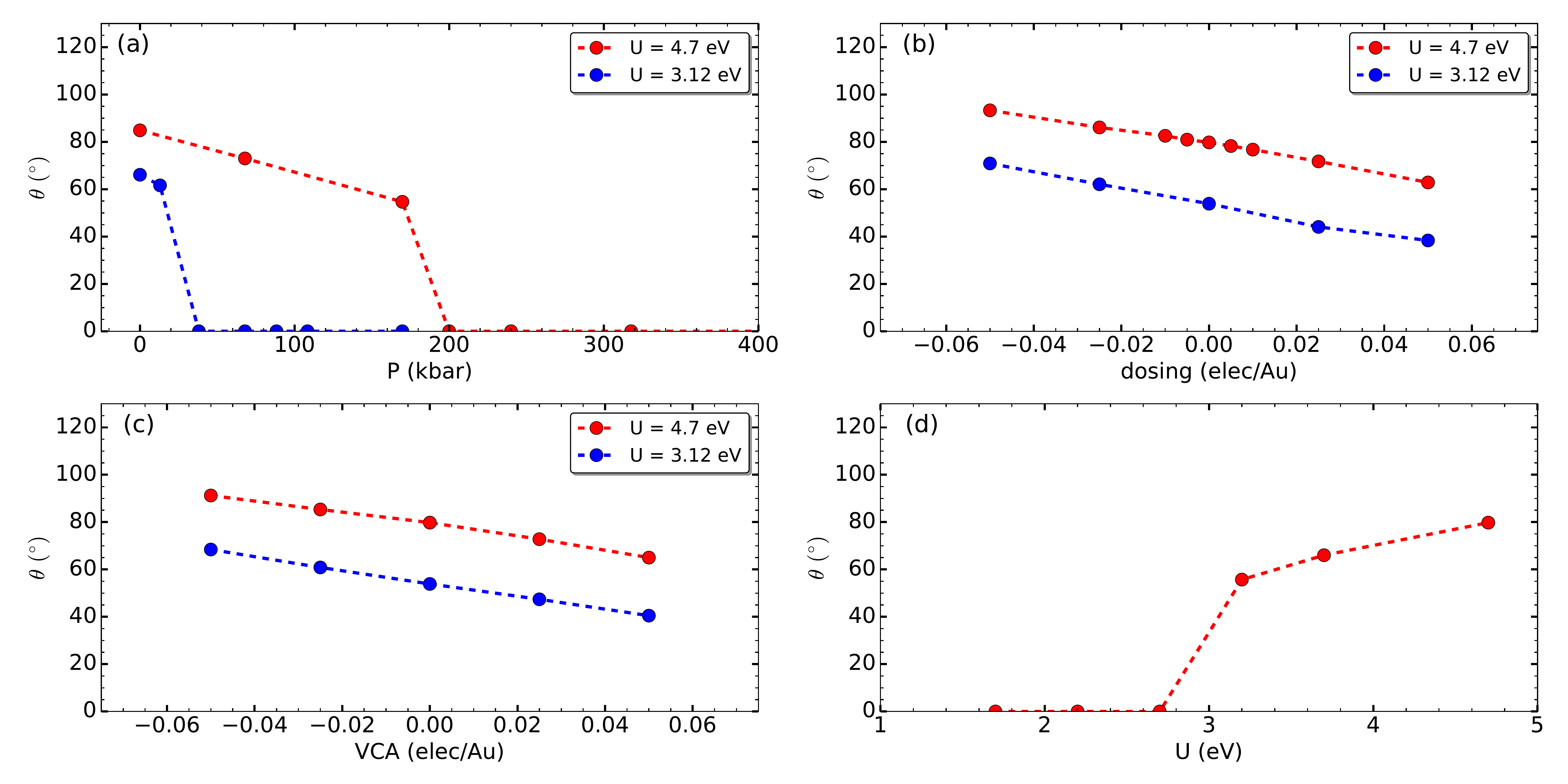}
  \caption{(Color online) The spiral angle $\theta$ as a function of (a)
    pressure, (b) charge dosing, (c) the VCA, and (d) $U$. Panels (a), (b), and
    (c) include results for $U = 3.12 \text{ eV}$ and $4.7 \text{ eV}$, see
    legend. The angle $\theta$ is obtained by minimizing the
    $J_{1}-J_{2}-J_{3}-J_{4}$ model.}
  \label{fig:angles}
\end{figure*}

\begin{figure}[t]
  \centering
  \includegraphics[width=0.49\textwidth]{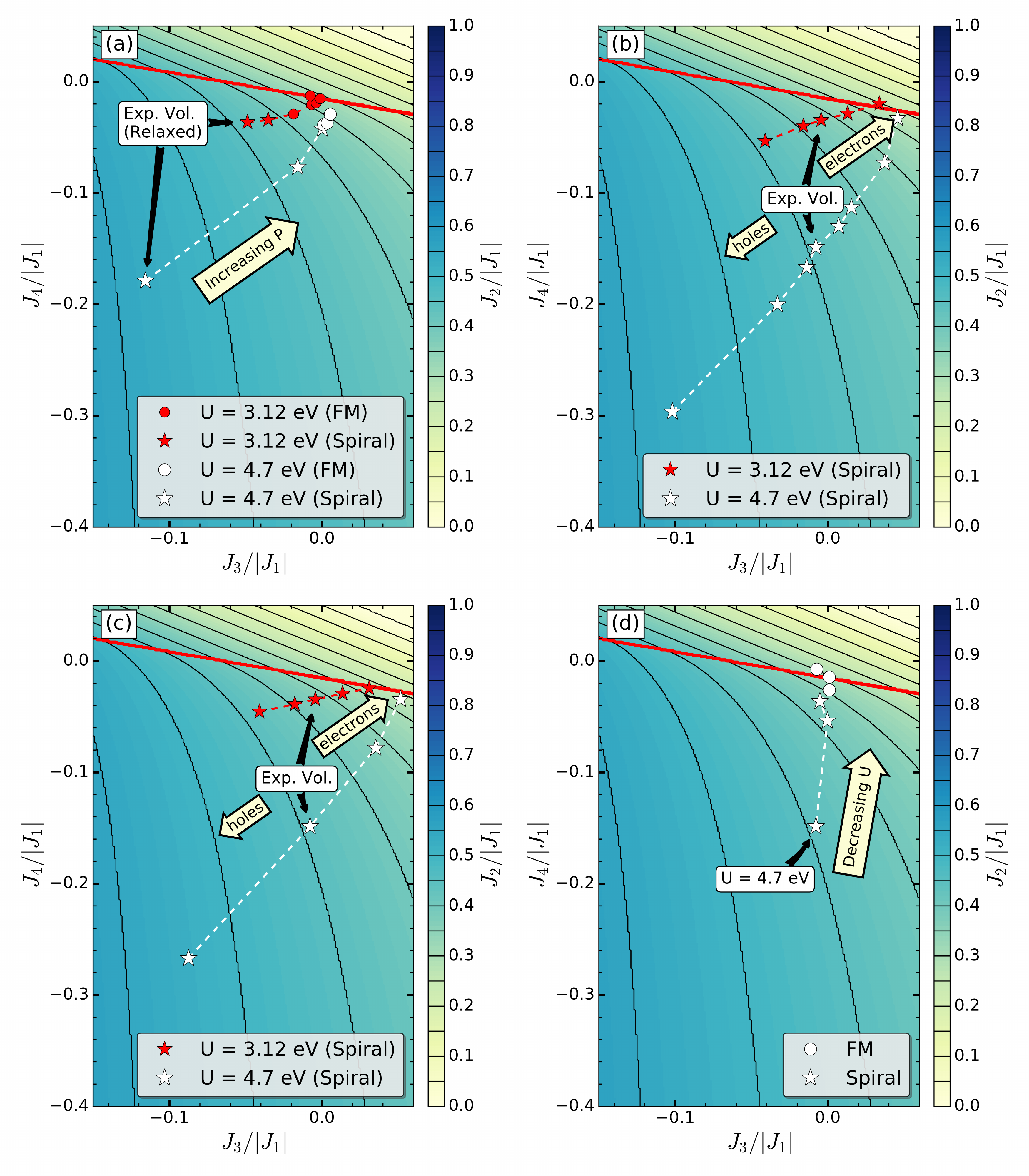}
  \caption{(Color online) Parametric plots of the MnAu$_{2}$ exchange parameters
    as a function of pressure, doping, and $U$. The parametric curves are
    visualized on top of the contour plot from
    Fig.~\ref{fig:heisenberg_model}(d). The plot marker shapes indicate the
    magnetic order of MnAu$_{2}$, with stars corresponding to spiral order and
    circles to ferromagnetic order. See legend for the values of $U$. (a)
    Parametric curve as a function of pressure. The reference point is the
    $P = 0$ unit cell, see Fig.~\ref{fig:lattice} for lattice parameters. The
    arrow indicates the direction of increasing pressure. (b) Parametric curve
    as a function of charge dosing. The reference point is a unit cell with
    experimental lattice parameters and no doping. The arrows indicate the
    direction for increasing electron/hole doping. (c) Parametric curve as a
    function of the VCA. The reference point is a unit cell with experimental
    lattice parameters and no doping. The arrows indicate the direction for
    increasing electron/hole doping. (d) Parametric curve as a function of
    $U$. The reference point is a unit cell with experimental parameters and
    $U = 4.7 \text{ eV}$. The arrow indicates the direction of decreasing $U$.}
  \label{fig:criticalparams}
\end{figure}

\begin{figure*}
  \centering
  \includegraphics[width=0.95\textwidth]{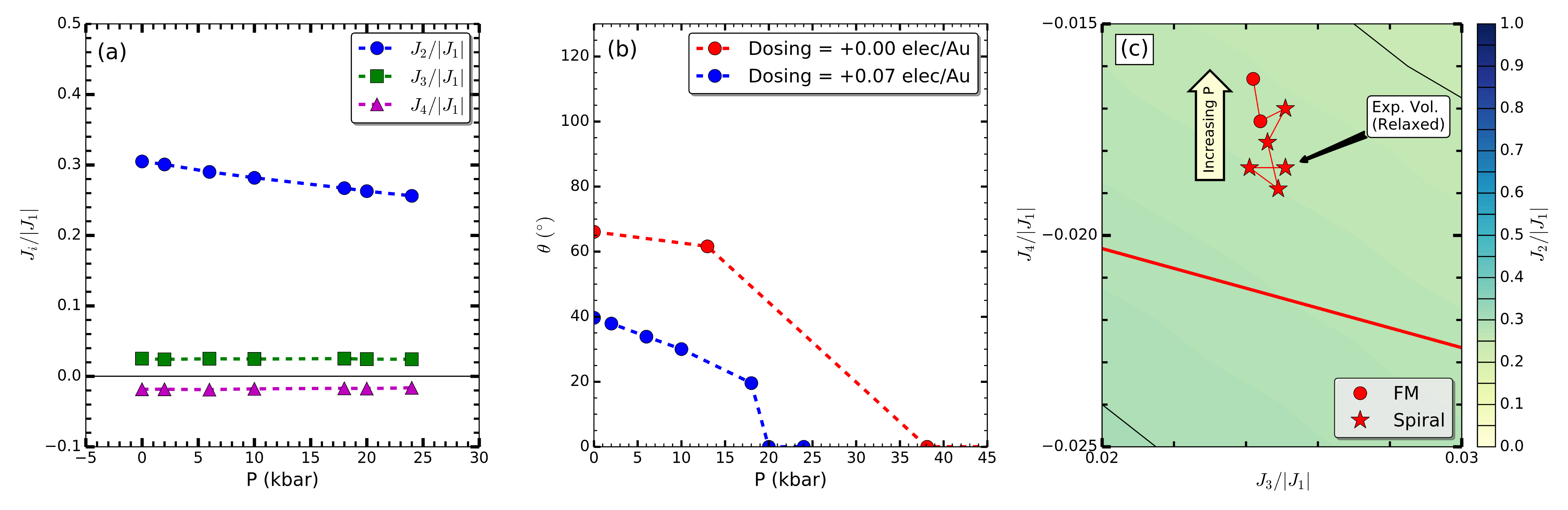}
  \caption{(Color online) Plots showing the results of calculations with
    simultaneous charge dosing and applied pressure. The calculations were
    performed using $U = 3.12 \text{ eV}$ and a charge dosing of $0.07$
    electrons per Au. (a) The exchange parameters $J_{2}/\abs{J_{1}}$,
    $J_{3}/\abs{J_{1}}$, and $J_{4}/\abs{J_{1}}$, see legend, as a function of
    pressure. (b) The spiral angle $\theta$ as a function of pressure with and
    without charge dosing, see the legend. (c) Parametric plot of the exchange
    parameters as a function of pressure visualized on top of the contour plot
    from Fig.~\ref{fig:heisenberg_model}(d). The plot markers indicate the type
    of magnetic order, see the legend. The reference point is the relaxed
    $P = 0$ unit cell. The arrow indicates the direction of increasing
    pressure.}
  \label{fig:charge_and_dosing}
\end{figure*}

It should be noted that these calculations and their subsequent analysis are all
done at zero temperature, a choice that we should justify. Measurements of the
saturation magnetization at low temperatures ($T \leq 10 \text{ K}$) and at room
temperature \cite{Nagata1999_JAlComp_Magnetism, Handstein2000_JAppl}, along with
measurements of the susceptibility and the critical magnetic field $H_{c}$
\cite{Nagata1999_JAlComp_Magnetism}, show a very weak dependence on
temperature. As discussed in Sec.~\ref{sec:introduction}, the temperature
dependence of the spiral angle as a function of pressure, obtained using neutron
diffraction measurements at 77 K and at room temperature, is also weak
\cite{Smith1966_JPhysChemSol_neutron, Handstein2005_JMag}. The weak temperature
dependence of these material parameters suggests that the internal energy
determines the important qualitative features of the magnetic phase diagram of
MnAu$_{2}$, and that including thermal fluctuations will result in only minor
changes to the area of the spiral and ferromagnetic regions and the length of
the first and second order phase boundaries. Therefore, it is reasonable to
expect that our conclusions about the spiral-to-ferromagnetic phase transition
obtained at zero temperature hold at finite temperatures.

For the pressure calculations, we constrained the volume of our cell and relaxed
the structure as described in Sec.~\ref{sec:comp-deta}, and then calculated the
pressure using the formula $P = -dE/dV$ (see Supplemental Material
\cite{supp_mater}). Using the LSDA functional during relaxation results in
overbinding, and so here the equilibrium volume is predicted to be 6\% smaller
than experiment. To account for this, we shift our pressures such that $P = 0$
corresponds to the experimental volume, and the lattice and internal parameters
as a function of pressure are shown in Fig.~\ref{fig:lattice}. As the pressure
is increased, the parameters decrease in a smooth way, which contrasts with the
experimental observation in Ref.~\onlinecite{Smith1966_JPhysChemSol_neutron},
which seemed to indicate a sharp drop in the c parameter for
$P > 6.5 \text{ kbar}$ \footnote{There is a large error in measuring $c$ as a
  function of pressure in Ref.~\onlinecite{Smith1966_JPhysChemSol_neutron}, so
  it is likely that the reported drop is not a real effect.}.

The fitted exchange parameters as functions of pressure, charge dosing, the VCA,
and $U$ are reported in Table \ref{tab:exchange_parameters}. We find that
$J_{1}$ becomes more ferromagnetic with increasing pressure, electron doping,
and decreasing $U$, and that it is also the most sensitive to these
parameters. At 28 kbar of pressure, the absolute value of $J_{1}$ grows by
$3.0 \text{ meV} \sim 35 \text{ K}$ for $U = 3.12$ eV and
$4.5 \text{ meV} \sim 52 \text{ K}$ for $U = 4.7$ eV. In
Ref.~\onlinecite{Wayne1969_JPhysChemSol_pressure}, the critical transition
temperature (N\'{e}el or Curie, depending on the magnetic state at the given
pressure) increases by $\sim 40 \text{ K}$ going from zero pressure to 28 kbar,
so these changes in $J_{1}$ are of the correct order.

To evaluate the spiral-to-ferromagnetic phase transition in the context of the
phase diagrams in Fig.~\ref{fig:heisenberg_model}, we plot the scaled exchange
parameters, $j_{i} \equiv J_{i} / \abs{J_{1}}$, in
Fig.~\ref{fig:exchangeparams}. We find that in all cases $j_{2}$ is always
antiferromagnetic and is the most sensitive to pressure, doping, and variation
in $U$. $j_{2}$ decreases with increasing pressure, electron doping, and
decreasing $U$. The exchange couplings $j_{3}$ and $j_{4}$ are closer to zero,
and do not respond to pressure, doping, and $U$ in the same way as
$j_{2}$. Applying pressure suppresses $j_{3}$, and for $U = 4.7 \text{ eV}$ the
coupling becomes slightly antiferromagnetic for $P \gtrsim 170 \text{
  kbar}$. Charge dosing and VCA doping have a stronger effect, where $j_{3}$
becomes more ferromagnetic with hole doping (for $U = 4.7 \text{ eV}$, the
coupling becomes enhanced at the large hole doping of -0.05 electrons per Au)
and turns antiferromagnetic with electron doping. However, $j_{3}$ is
insensitive to $U$, staying close to zero for all considered values. $j_{4}$,
unlike $j_{3}$, remains ferromagnetic in all cases, and often
$\abs{j_{4}} > \abs{j_{3}}$. Pressure affects $j_{4}$ in a similar way as
$j_{3}$, and for $U = 4.7 \text{ eV}$ so does doping, but for
$U = 3.12 \text{ eV}$, $j_{4}$ is less sensitive to doping than
$j_{3}$. Finally, decreasing $U$ suppresses $j_{4}$, going from moderate
ferromagnetic coupling at $U = 4.7 \text{ eV}$ to being nearly zero at
$U = 1.2 \text{ eV}$. This indicates that measuring the fourth neighbor
interplanar coupling in MnAu$_{2}$ would yield information about the strength of
the electronic correlations in this material.

Next we plot $\theta$ as a function of pressure, charge dosing, VCA doping, and
$U$ in Fig.~\ref{fig:angles}. We obtain $\theta$ by minimizing Eq.~\eqref{eq:2}
with the fitted exchange parameters \footnote{We note that an alternative
  approach is to interpolate $E(\theta)$ and find the minimum, and away from
  phase transitions these two methods are in good agreement (see Supplemental
  Material \cite{supp_mater}). However, near a phase transition the energy wells
  become shallow and the interpolations are no longer reliable, so for
  consistency we report results obtained by minimizing Eq.~\eqref{eq:2}}. We
find that $\theta$ is quite sensitive to pressure, doping, and $U$. As expected,
applying pressure decreases $\theta$ in agreement with
Ref.~\onlinecite{Smith1966_JPhysChemSol_neutron}, and at a critical pressure
this induces a phase transition, which is in the range
$12.977 \leq P \leq 38.109$ kbar for $U = 3.12 \text{ eV}$ and
$169.709 \leq P \leq 200$ kbar for $U = 4.7 \text{ eV}$. This indicates that
measurements of the critical pressure provides information about the strength of
the electronic correlations. In addition, for both charge dosing and VCA doping
the rate of change of $\theta$ is independent of $U$, and $\theta$ decreases
with electron doping and increases with hole doping. Finally, in
Fig.~\ref{fig:angles}(d) we find that $\theta$ decreases with decreasing $U$,
and eventually collapses to a ferromagnetic state.

The evolution of the magnetic coupling of MnAu$_{2}$ when subjected to pressure,
doping, and variation in $U$ can be further visualized using the contour plot in
Fig.~\ref{fig:heisenberg_model}(d). We take the exchange parameters $j_{3}$ and
$j_{4}$ and plot them as parametric curves on top of the contour plot, which
captures how they vary with pressure, doping, and $U$. For each point, we also
compare the fitted $j_{2}$ with the critical $j_{2}^{c}$ that marks the
spiral-to-ferromagnetic phase transition. If $j_{2} < j_{2}^{c}$, we plot the
point as a circle, indicating that the system is ferromagnetic, and if
$j_{2} \geq j_{2}^{c}$ we plot the point as a star, which represents spiral
order. This procedure yields Fig.~\ref{fig:criticalparams}, and on each panel we
give a reference point, which for panel (a) corresponds to the relaxed cell at
experimental volume ($P = 0$), for panels (b) and (c) corresponds to the
experimental lattice parameters with zero doping, and for panel (d) corresponds
to experimental lattice parameters with $U = 4.7 \text{ eV}$. The arrows with
text indicate the direction that the parametric curves evolve, so following the
curve from the reference point in panel (a) shows how the parameters change with
increasing pressure, in panels (b) and (c) one direction is electron doping and
the other is hole doping, and for panel (d) $U$ decreases along the curve. If,
while following the curve, the marker symbol changes from a star to a circle,
this corresponds to a phase transition. If this change in symbols happens while
south of the red line, then the phase transition is first order, and if it
happens north of the red line, then it is a second order.

The parametric path for MnAu$_{2}$ under pressure in
Fig.~\ref{fig:criticalparams}(a) shows that the spiral-to-ferromagnetic
transition is first order for both values of $U$. So, for ideal, stoichiometric
MnAu$_{2}$ samples, we expect the pressure-induced phase transition to be of
first order, in agreement with
Ref.~\onlinecite{Adiatullin1971_SovPhysSolSt_Influence}. Decreasing $U$ at
ambient pressure also induces a first order phase transition. In contrast, for
our considered doping levels MnAu$_{2}$ does not collapse to the ferromagnetic
state. Doping with holes stabilizes the spiral state and drives the system
deeper into the first-order parameter region, while doping with electrons moves
MnAu$_{2}$ closer to ferromagnetism and also towards a second-order phase
transition. Also, the stronger the correlations, the more dramatic the shifts in
parameter space with doping. For both $U = 3.12 \text{ eV}$ and $4.7 \text{ eV}$
an electron doping of 0.05 electrons per Au places the system at the dividing
line between first and second order, and it would be expected that the
application of pressure for this system would induce a second order transition.

\begin{table}[b]
  \centering
  \begin{tabular}[c]{c c c c c}
    \toprule
    Pressure & $J_{1}$ & $J_{2}$ & $J_{3}$ & $J_{4}$ \\
    (kbar) & \multicolumn{4}{c}{(meV)} \\ \midrule
     0.000 & -33.5978 & 10.2441 & 0.8442 & -0.6193 \\
     2.000 & -33.9957 & 10.2223 & 0.8197 & -0.6252 \\
     6.000 & -34.8238 & 10.1037 & 0.8679 & -0.6570 \\
    10.000 & -35.6126 & 10.0322 & 0.8756 & -0.6331 \\
    18.000 & -37.0886 &  9.9025 & 0.9327 & -0.6302 \\
    20.000 & -37.4883 &  9.8482 & 0.9131 & -0.6498 \\
    24.000 & -38.1828 &  9.7781 & 0.9236 & -0.6241 \\
    \bottomrule
  \end{tabular}
  \caption{Values of the fitted exchange parameters for the simultaneous
    pressure and charge dosing calculations. Results are for a charge dosing
    value of 0.07 electrons per Au and $U = 3.12 \text{ eV}$.}
  \label{tab:charge_pressure_exchange_parameters}
\end{table}

We check the prediction of inducing a second order phase transition by
calculating the pressure dependence of MnAu$_{2}$ with $U = 3.12 \text{ eV}$ and
a charge dosing level of 0.07 electrons per Au. The exchange parameters obtained
from these calculations are reported in Table
\ref{tab:charge_pressure_exchange_parameters} and the scaled exchange
parameters, pressure dependence of $\theta$, and a parametric plot of the
exchange parameters are plotted in Fig.~\ref{fig:charge_and_dosing}. As shown in
Fig.~\ref{fig:charge_and_dosing}(a) the initial charge dosing biases $j_{2}$ to
be less antiferromagnetic and for $j_{3}$ to become more antiferromagnetic. For
the range of pressures considered, $j_{2}$ decreases with increasing pressure
while $j_{3}$ and $j_{4}$ are unaffected. Fig.~\ref{fig:charge_and_dosing}(b)
shows the effect this has on the $\theta$ vs pressure curve, where the trend of
the charge-dosed curve is consistent with a second order phase transition,
particularly when compared with the first order transition of the undoped curve
included for comparison. The parametric curve in
Fig.~\ref{fig:charge_and_dosing}(c) shows that the charge-dosed
pressure-dependent path in exchange parameter space is north of the dividing
line, confirming the prediction that applying pressure to electron doped
MnAu$_{2}$ can induce a second-order phase transition.

We can use the result that applying pressure to electron-doped MnAu$_{2}$
induces a second order phase transition to resolve the contradiction in the
pressure experiments. In Ref.~\onlinecite{Wayne1969_JPhysChemSol_pressure},
where the authors found a second order phase transition for MnAu$_{2}$, they
reported that there were trace amounts of MnAu in their sample, such that the
system was 34.3\% Mn and 65.7\% Au. This excess of Mn in their sample can lead
to an effective electron doping. Let us assume that the excess Mn atoms are also
in the $2+$ ionic charge state, and that the extra electrons they donate are
spread homogeneously over the system. This would mean that each unit cell of
MnAu$_{2}$ would have
$2 \text{ electrons per Mn} \cdot (34.3/33.3) = 2.06 \text{ electrons per Mn}$,
or an excess of 0.03 electrons per Au.  Assuming that $U = 3.12 \text{ eV}$
\footnote{This choice of $U$ yields a $\theta < 60^{\circ}$ and a critical
  transition pressure $P \sim 20-40 \text{ kbar}$, in reasonable agreement with
  experiment.} and comparing with the electron dosing curve in
Fig.~\ref{fig:criticalparams}(b), an extra 0.03 electrons per Au would shift the
sample near the second order transition region of the parameter space, which as
we've seen can lead to a second order phase transition upon application of
pressure. While the MnAu domains are not likely to uniformly dope the material,
we see how the variations in sample quality can influence the kind of phase
transition. This both stresses the importance of controlling for impurities in
MnAu$_{2}$, and that the magnetic properties are also amenable to tuning.

\section{Conclusions}
\label{sec:conclusions}

We studied how the magnetic exchange in MnAu$_{2}$ changes upon application of
pressure and doping using a series of DFT calculations, as well as varying the
electronic correlations with a Hubbard-like $U$. We found that the proper model
for these interactions is the one-dimensional Heisenberg model with coupling
between the first \textit{four} Mn planes (the $J_{1} - J_{2} - J_{3} - J_{4}$
model), which contrasts with previous studies that only considered interactions
between the first two planes. Within this model, transitions between the
ferromagnetic and spiral phases can be either first or second order, and $J_{3}$
and $J_{4}$ control the order of the phase transition. Our analysis finds that
ideal, stoichiometric MnAu$_{2}$ sits near a dividing line in $(J_{3}, J_{4})$
space that separates first- and second-order phase transitions. Applying
pressure collapses the spiral state as expected, and for ideal, stoichiometric
MnAu$_{2}$ the transition is first order. Hole doping the material further
stabilizes the spiral and the tendency towards a first-order transition. In
contrast, electron doping the material makes the spiral less stable and moves
the system in $(J_{3}, J_{4})$ space closer to a second-order transition. We
confirm with additional DFT calculations that applying pressure to
electron-doped MnAu$_{2}$ will induce a second-order phase
transition. Therefore, impurities that provide an effective electron doping,
such as excess Mn, can cause the spiral-to-ferromagnetic phase transition to
become second order under pressure. These results indicate that MnAu$_{2}$ is
tunable, where the kind of phase transition can be set and the spiral angle
controlled through both pressure and doping, which can include gate
voltages. The ability to control the spiral angle holds promise for integration
of MnAu$_{2}$ in spintronics devices, such as in a spin valve. Attaching a thin
film of MnAu$_{2}$ to the free magnetic layer of a spin valve and then applying
a gate voltage could allow for a current-free manipulation of the relative
orientation between the free and pinned magnetic layers, which would reduce
Joule heating and could help in developing smaller-scale electronics devices.

\begin{acknowledgments}
  J.K.G.~thanks Igor Mazin and Konrad Bussmann for useful discussions regarding
  this project, and acknowledges the support of the NRC program at NRL.
\end{acknowledgments}

\appendix

\section*{Supplementary Materials}
\label{sec:suppl-mater}

The supplemental materials for this report are provided in the IPython notebook
format. The notebook file and the data used for analysis are provided at
\url{https://github.com/jkglasbrenner/ipython_nb_collapse_and_control_of_the_mnau2_spin_spiral_state_through_pressure_and_doping}. For
a readable HTML version of the notebook, download the following file and open it
in your web-browser:
\url{https://github.com/jkglasbrenner/ipython_nb_collapse_and_control_of_the_mnau2_spin_spiral_state_through_pressure_and_doping/raw/master/ipython_nb_MnAu2_pressure_supplemental_material.html}


\begin{thebibliography}{39}%
\makeatletter
\providecommand \@ifxundefined [1]{%
 \@ifx{#1\undefined}
}%
\providecommand \@ifnum [1]{%
 \ifnum #1\expandafter \@firstoftwo
 \else \expandafter \@secondoftwo
 \fi
}%
\providecommand \@ifx [1]{%
 \ifx #1\expandafter \@firstoftwo
 \else \expandafter \@secondoftwo
 \fi
}%
\providecommand \natexlab [1]{#1}%
\providecommand \enquote  [1]{``#1''}%
\providecommand \bibnamefont  [1]{#1}%
\providecommand \bibfnamefont [1]{#1}%
\providecommand \citenamefont [1]{#1}%
\providecommand \href@noop [0]{\@secondoftwo}%
\providecommand \href [0]{\begingroup \@sanitize@url \@href}%
\providecommand \@href[1]{\@@startlink{#1}\@@href}%
\providecommand \@@href[1]{\endgroup#1\@@endlink}%
\providecommand \@sanitize@url [0]{\catcode `\\12\catcode `\$12\catcode
  `\&12\catcode `\#12\catcode `\^12\catcode `\_12\catcode `\%12\relax}%
\providecommand \@@startlink[1]{}%
\providecommand \@@endlink[0]{}%
\providecommand \url  [0]{\begingroup\@sanitize@url \@url }%
\providecommand \@url [1]{\endgroup\@href {#1}{\urlprefix }}%
\providecommand \urlprefix  [0]{URL }%
\providecommand \Eprint [0]{\href }%
\providecommand \doibase [0]{http://dx.doi.org/}%
\providecommand \selectlanguage [0]{\@gobble}%
\providecommand \bibinfo  [0]{\@secondoftwo}%
\providecommand \bibfield  [0]{\@secondoftwo}%
\providecommand \translation [1]{[#1]}%
\providecommand \BibitemOpen [0]{}%
\providecommand \bibitemStop [0]{}%
\providecommand \bibitemNoStop [0]{.\EOS\space}%
\providecommand \EOS [0]{\spacefactor3000\relax}%
\providecommand \BibitemShut  [1]{\csname bibitem#1\endcsname}%
\let\auto@bib@innerbib\@empty
%</preamble>
\bibitem [{\citenamefont {Meyer}\ and\ \citenamefont
  {{Taglang}}(1956)}]{Meyer1956_JPhysRadium_Proprietes}%
  \BibitemOpen
  \bibfield  {author} {\bibinfo {author} {\bibfnamefont {A.~J.}\ \bibnamefont
  {Meyer}}\ and\ \bibinfo {author} {\bibfnamefont {P.}~\bibnamefont
  {{Taglang}}},\ }\href@noop {} {\bibfield  {journal} {\bibinfo  {journal}
  {J.~Phys.~Radium}\ }\textbf {\bibinfo {volume} {17}},\ \bibinfo {pages} {457}
  (\bibinfo {year} {1956})}\BibitemShut {NoStop}%
\bibitem [{\citenamefont {Herpin}\ \emph {et~al.}(1959)\citenamefont {Herpin},
  \citenamefont {{Meriel}},\ and\ \citenamefont
  {{Villain}}}]{Herpin1959_ComptesRendus_Structure}%
  \BibitemOpen
  \bibfield  {author} {\bibinfo {author} {\bibfnamefont {A.}~\bibnamefont
  {Herpin}}, \bibinfo {author} {\bibfnamefont {P.}~\bibnamefont {{Meriel}}}, \
  and\ \bibinfo {author} {\bibfnamefont {J.}~\bibnamefont {{Villain}}},\
  }\href@noop {} {\bibfield  {journal} {\bibinfo  {journal} {Comptes Rendus}\
  }\textbf {\bibinfo {volume} {249}},\ \bibinfo {pages} {1334} (\bibinfo {year}
  {1959})}\BibitemShut {NoStop}%
\bibitem [{\citenamefont {Herpin}\ and\ \citenamefont
  {{Meriel}}(1961)}]{Herpin1961_JPhysRadium_Etude}%
  \BibitemOpen
  \bibfield  {author} {\bibinfo {author} {\bibfnamefont {A.}~\bibnamefont
  {Herpin}}\ and\ \bibinfo {author} {\bibfnamefont {P.}~\bibnamefont
  {{Meriel}}},\ }\href@noop {} {\bibfield  {journal} {\bibinfo  {journal}
  {J.~Phys.~Radium}\ }\textbf {\bibinfo {volume} {22}},\ \bibinfo {pages} {337}
  (\bibinfo {year} {1961})}\BibitemShut {NoStop}%
\bibitem [{\citenamefont {Samata}\ \emph {et~al.}(1998)\citenamefont {Samata},
  \citenamefont {{Sekiguchi}}, \citenamefont {{Sawabe}}, \citenamefont
  {{Nagata}}, \citenamefont {{Uchida}},\ and\ \citenamefont
  {{Lan}}}]{Samata1998_JPhysChemSol_Giant}%
  \BibitemOpen
  \bibfield  {author} {\bibinfo {author} {\bibfnamefont {H.}~\bibnamefont
  {Samata}}, \bibinfo {author} {\bibfnamefont {N.}~\bibnamefont {{Sekiguchi}}},
  \bibinfo {author} {\bibfnamefont {A.}~\bibnamefont {{Sawabe}}}, \bibinfo
  {author} {\bibfnamefont {Y.}~\bibnamefont {{Nagata}}}, \bibinfo {author}
  {\bibfnamefont {T.}~\bibnamefont {{Uchida}}}, \ and\ \bibinfo {author}
  {\bibfnamefont {M.~D.}\ \bibnamefont {{Lan}}},\ }\href@noop {} {\bibfield
  {journal} {\bibinfo  {journal} {J.~Phys.~Chem.~Sol.}\ }\textbf {\bibinfo
  {volume} {59}},\ \bibinfo {pages} {377} (\bibinfo {year} {1998})}\BibitemShut
  {NoStop}%
\bibitem [{\citenamefont {Smith}\ \emph {et~al.}(1966)\citenamefont {Smith},
  \citenamefont {{Bradley}},\ and\ \citenamefont
  {{Bacon}}}]{Smith1966_JPhysChemSol_neutron}%
  \BibitemOpen
  \bibfield  {author} {\bibinfo {author} {\bibfnamefont {F.~A.}\ \bibnamefont
  {Smith}}, \bibinfo {author} {\bibfnamefont {C.~C.}\ \bibnamefont
  {{Bradley}}}, \ and\ \bibinfo {author} {\bibfnamefont {G.~E.}\ \bibnamefont
  {{Bacon}}},\ }\href@noop {} {\bibfield  {journal} {\bibinfo  {journal}
  {J.~Phys.~Chem.~Sold.}\ }\textbf {\bibinfo {volume} {27}},\ \bibinfo {pages}
  {925} (\bibinfo {year} {1966})}\BibitemShut {NoStop}%
\bibitem [{\citenamefont {{Handstein}}\ \emph {et~al.}(2005)\citenamefont
  {{Handstein}}, \citenamefont {{R{\"o}{\ss}ler}}, \citenamefont
  {{Idzikowski}}, \citenamefont {{Kozlova}}, \citenamefont {{Nenkov}},
  \citenamefont {{M{\"u}ller}}, \citenamefont {{Kreyssig}}, \citenamefont
  {{Loewenhaupt}}, \citenamefont {{Heinemann}}, \citenamefont {{Hoell}},\ and\
  \citenamefont {{St{\"u}{\ss}er}}}]{Handstein2005_JMag}%
  \BibitemOpen
  \bibfield  {author} {\bibinfo {author} {\bibfnamefont {A.}~\bibnamefont
  {{Handstein}}}, \bibinfo {author} {\bibfnamefont {U.~K.}\ \bibnamefont
  {{R{\"o}{\ss}ler}}}, \bibinfo {author} {\bibfnamefont {B.}~\bibnamefont
  {{Idzikowski}}}, \bibinfo {author} {\bibfnamefont {N.}~\bibnamefont
  {{Kozlova}}}, \bibinfo {author} {\bibfnamefont {K.}~\bibnamefont {{Nenkov}}},
  \bibinfo {author} {\bibfnamefont {K.~H.}\ \bibnamefont {{M{\"u}ller}}},
  \bibinfo {author} {\bibfnamefont {A.}~\bibnamefont {{Kreyssig}}}, \bibinfo
  {author} {\bibfnamefont {M.}~\bibnamefont {{Loewenhaupt}}}, \bibinfo {author}
  {\bibfnamefont {A.}~\bibnamefont {{Heinemann}}}, \bibinfo {author}
  {\bibfnamefont {A.}~\bibnamefont {{Hoell}}}, \ and\ \bibinfo {author}
  {\bibfnamefont {N.}~\bibnamefont {{St{\"u}{\ss}er}}},\ }\href@noop {}
  {\bibfield  {journal} {\bibinfo  {journal} {J.~Mag.~Mag.~Mater.}\ }\textbf
  {\bibinfo {volume} {290-291, Part 2}},\ \bibinfo {pages} {1093} (\bibinfo
  {year} {2005})}\BibitemShut {NoStop}%
\bibitem [{\citenamefont {{Evenson}}\ and\ \citenamefont
  {{Liu}}(1968)}]{Evenson1968_PhysRevLett_Generalized}%
  \BibitemOpen
  \bibfield  {author} {\bibinfo {author} {\bibfnamefont {W.~E.}\ \bibnamefont
  {{Evenson}}}\ and\ \bibinfo {author} {\bibfnamefont {S.~H.}\ \bibnamefont
  {{Liu}}},\ }\href@noop {} {\bibfield  {journal} {\bibinfo  {journal}
  {Phys.~Rev.~Lett.}\ }\textbf {\bibinfo {volume} {21}},\ \bibinfo {pages}
  {432} (\bibinfo {year} {1968})}\BibitemShut {NoStop}%
\bibitem [{\citenamefont {{Ruderman}}\ and\ \citenamefont
  {{Kittel}}(1954)}]{Ruderman1954_PhysRev_Indirect}%
  \BibitemOpen
  \bibfield  {author} {\bibinfo {author} {\bibfnamefont {M.~A.}\ \bibnamefont
  {{Ruderman}}}\ and\ \bibinfo {author} {\bibfnamefont {C.}~\bibnamefont
  {{Kittel}}},\ }\href@noop {} {\bibfield  {journal} {\bibinfo  {journal}
  {Phys.~Rev.}\ }\textbf {\bibinfo {volume} {96}},\ \bibinfo {pages} {99}
  (\bibinfo {year} {1954})}\BibitemShut {NoStop}%
\bibitem [{\citenamefont {{Kasuya}}(1956)}]{Kasuya1956_ProgTheorPhys_Theory}%
  \BibitemOpen
  \bibfield  {author} {\bibinfo {author} {\bibfnamefont {T.}~\bibnamefont
  {{Kasuya}}},\ }\href@noop {} {\bibfield  {journal} {\bibinfo  {journal}
  {Prog.~Theor.~Phys.}\ }\textbf {\bibinfo {volume} {16}},\ \bibinfo {pages}
  {45} (\bibinfo {year} {1956})}\BibitemShut {NoStop}%
\bibitem [{\citenamefont {{Yosida}}(1957)}]{Yosida1957_PhysRev_Magnetic}%
  \BibitemOpen
  \bibfield  {author} {\bibinfo {author} {\bibfnamefont {K.}~\bibnamefont
  {{Yosida}}},\ }\href@noop {} {\bibfield  {journal} {\bibinfo  {journal}
  {Phys.~Rev.}\ }\textbf {\bibinfo {volume} {106}},\ \bibinfo {pages} {893}
  (\bibinfo {year} {1957})}\BibitemShut {NoStop}%
\bibitem [{\citenamefont
  {{Dzyaloshinsky}}(1958)}]{Dzyaloshinsky1958_JPhysChemSol_thermodynamic}%
  \BibitemOpen
  \bibfield  {author} {\bibinfo {author} {\bibfnamefont {I.}~\bibnamefont
  {{Dzyaloshinsky}}},\ }\href@noop {} {\bibfield  {journal} {\bibinfo
  {journal} {J.~Phys.~Chem.~Sol.}\ }\textbf {\bibinfo {volume} {4}},\ \bibinfo
  {pages} {241} (\bibinfo {year} {1958})}\BibitemShut {NoStop}%
\bibitem [{\citenamefont {{Moriya}}(1960)}]{Moriya1960_PhysRev_Anisotropic}%
  \BibitemOpen
  \bibfield  {author} {\bibinfo {author} {\bibfnamefont {T.}~\bibnamefont
  {{Moriya}}},\ }\href@noop {} {\bibfield  {journal} {\bibinfo  {journal}
  {Phys.~Rev.}\ }\textbf {\bibinfo {volume} {120}},\ \bibinfo {pages} {91}
  (\bibinfo {year} {1960})}\BibitemShut {NoStop}%
\bibitem [{\citenamefont {{Bak}}\ and\ \citenamefont
  {{Jensen}}(1980)}]{Bak1980_JPhysC_Theory}%
  \BibitemOpen
  \bibfield  {author} {\bibinfo {author} {\bibfnamefont {P.}~\bibnamefont
  {{Bak}}}\ and\ \bibinfo {author} {\bibfnamefont {M.~H.}\ \bibnamefont
  {{Jensen}}},\ }\href@noop {} {\bibfield  {journal} {\bibinfo  {journal}
  {J.~Phys.~C}\ }\textbf {\bibinfo {volume} {13}},\ \bibinfo {pages} {L881}
  (\bibinfo {year} {1980})}\BibitemShut {NoStop}%
\bibitem [{\citenamefont {{Nakanishi}}\ \emph {et~al.}(1980)\citenamefont
  {{Nakanishi}}, \citenamefont {{Yanase}}, \citenamefont {{Hasegawa}},\ and\
  \citenamefont {{Kataoka}}}]{Nakanishi1980_SolStComm_origin}%
  \BibitemOpen
  \bibfield  {author} {\bibinfo {author} {\bibfnamefont {O.}~\bibnamefont
  {{Nakanishi}}}, \bibinfo {author} {\bibfnamefont {A.}~\bibnamefont
  {{Yanase}}}, \bibinfo {author} {\bibfnamefont {A.}~\bibnamefont
  {{Hasegawa}}}, \ and\ \bibinfo {author} {\bibfnamefont {M.}~\bibnamefont
  {{Kataoka}}},\ }\href@noop {} {\bibfield  {journal} {\bibinfo  {journal}
  {Sol.~St.~Comm.}\ }\textbf {\bibinfo {volume} {35}},\ \bibinfo {pages} {995}
  (\bibinfo {year} {1980})}\BibitemShut {NoStop}%
\bibitem [{\citenamefont {{Kataoka}}\ and\ \citenamefont
  {{Nakanishi}}(1981)}]{Kataoka1981_JPhysSocJpn_Helical}%
  \BibitemOpen
  \bibfield  {author} {\bibinfo {author} {\bibfnamefont {M.}~\bibnamefont
  {{Kataoka}}}\ and\ \bibinfo {author} {\bibfnamefont {O.}~\bibnamefont
  {{Nakanishi}}},\ }\href@noop {} {\bibfield  {journal} {\bibinfo  {journal}
  {J.~Phys.~Soc.~Jpn.}\ }\textbf {\bibinfo {volume} {50}},\ \bibinfo {pages}
  {3888} (\bibinfo {year} {1981})}\BibitemShut {NoStop}%
\bibitem [{\citenamefont {Enz}(1961)}]{Enz1961_JApplPhys_Magnetization}%
  \BibitemOpen
  \bibfield  {author} {\bibinfo {author} {\bibfnamefont {U.}~\bibnamefont
  {Enz}},\ }\href@noop {} {\bibfield  {journal} {\bibinfo  {journal}
  {J.~Appl.~Phys.}\ }\textbf {\bibinfo {volume} {32}},\ \bibinfo {pages} {S22}
  (\bibinfo {year} {1961})}\BibitemShut {NoStop}%
\bibitem [{\citenamefont
  {Nagamiya}(1962)}]{Nagamiya1962_JApplPhys_Modification}%
  \BibitemOpen
  \bibfield  {author} {\bibinfo {author} {\bibfnamefont {T.}~\bibnamefont
  {Nagamiya}},\ }\href@noop {} {\bibfield  {journal} {\bibinfo  {journal}
  {J.~Appl.~Phys.}\ }\textbf {\bibinfo {volume} {33}},\ \bibinfo {pages} {1029}
  (\bibinfo {year} {1962})}\BibitemShut {NoStop}%
\bibitem [{\citenamefont {Udvardi}\ \emph {et~al.}(2006)\citenamefont
  {Udvardi}, \citenamefont {{Khmelevskyi}}, \citenamefont {{Szunyogh}},
  \citenamefont {{Mohn}},\ and\ \citenamefont
  {{Weinberger}}}]{Udvardi2006_PhysRevB_Helimagnetism}%
  \BibitemOpen
  \bibfield  {author} {\bibinfo {author} {\bibfnamefont {L.}~\bibnamefont
  {Udvardi}}, \bibinfo {author} {\bibfnamefont {S.}~\bibnamefont
  {{Khmelevskyi}}}, \bibinfo {author} {\bibfnamefont {L.}~\bibnamefont
  {{Szunyogh}}}, \bibinfo {author} {\bibfnamefont {P.}~\bibnamefont {{Mohn}}},
  \ and\ \bibinfo {author} {\bibfnamefont {P.}~\bibnamefont {{Weinberger}}},\
  }\href@noop {} {\bibfield  {journal} {\bibinfo  {journal} {Phys.~Rev.~B}\
  }\textbf {\bibinfo {volume} {73}},\ \bibinfo {pages} {104446} (\bibinfo
  {year} {2006})}\BibitemShut {NoStop}%
\bibitem [{\citenamefont {Glasbrenner}\ \emph {et~al.}(2014)\citenamefont
  {Glasbrenner}, \citenamefont {{Bussmann}},\ and\ \citenamefont
  {{Mazin}}}]{Glasbrenner2014_PhysRevB_Magnetic}%
  \BibitemOpen
  \bibfield  {author} {\bibinfo {author} {\bibfnamefont {J.~K.}\ \bibnamefont
  {Glasbrenner}}, \bibinfo {author} {\bibfnamefont {K.~M.}\ \bibnamefont
  {{Bussmann}}}, \ and\ \bibinfo {author} {\bibfnamefont {I.~I.}\ \bibnamefont
  {{Mazin}}},\ }\href@noop {} {\bibfield  {journal} {\bibinfo  {journal}
  {Phys.~Rev.~B}\ }\textbf {\bibinfo {volume} {90}},\ \bibinfo {pages} {144421}
  (\bibinfo {year} {2014})}\BibitemShut {NoStop}%
\bibitem [{\citenamefont {Grazhdankina}\ and\ \citenamefont
  {{Rodionov}}(1963)}]{Grazhdankina1963_SovPhysJETP_Effect}%
  \BibitemOpen
  \bibfield  {author} {\bibinfo {author} {\bibfnamefont {N.~P.}\ \bibnamefont
  {Grazhdankina}}\ and\ \bibinfo {author} {\bibfnamefont {K.~P.}\ \bibnamefont
  {{Rodionov}}},\ }\href@noop {} {\bibfield  {journal} {\bibinfo  {journal}
  {Sov.~Phys.~JETP}\ }\textbf {\bibinfo {volume} {16}},\ \bibinfo {pages}
  {1429} (\bibinfo {year} {1963})}\BibitemShut {NoStop}%
\bibitem [{\citenamefont {Wayne}\ and\ \citenamefont
  {{Smith}}(1969)}]{Wayne1969_JPhysChemSol_pressure}%
  \BibitemOpen
  \bibfield  {author} {\bibinfo {author} {\bibfnamefont {R.~C.}\ \bibnamefont
  {Wayne}}\ and\ \bibinfo {author} {\bibfnamefont {F.~A.}\ \bibnamefont
  {{Smith}}},\ }\href@noop {} {\bibfield  {journal} {\bibinfo  {journal}
  {J.~Phys.~Chem.~Sol.}\ }\textbf {\bibinfo {volume} {30}},\ \bibinfo {pages}
  {183} (\bibinfo {year} {1969})}\BibitemShut {NoStop}%
\bibitem [{\citenamefont {Adiatullin}\ and\ \citenamefont
  {{Fakidov}}(1971)}]{Adiatullin1971_SovPhysSolSt_Influence}%
  \BibitemOpen
  \bibfield  {author} {\bibinfo {author} {\bibfnamefont {R.}~\bibnamefont
  {Adiatullin}}\ and\ \bibinfo {author} {\bibfnamefont {I.~G.}\ \bibnamefont
  {{Fakidov}}},\ }\href@noop {} {\bibfield  {journal} {\bibinfo  {journal}
  {Sov.~Phys.~Sol.~St.}\ }\textbf {\bibinfo {volume} {12}},\ \bibinfo {pages}
  {2547} (\bibinfo {year} {1971})}\BibitemShut {NoStop}%
\bibitem [{\citenamefont {Pimpinelli}\ \emph {et~al.}(1989)\citenamefont
  {Pimpinelli}, \citenamefont {{Rastelli}},\ and\ \citenamefont
  {{Tassi}}}]{Pimpinelli1989_JPhysCondensMatter_Classical}%
  \BibitemOpen
  \bibfield  {author} {\bibinfo {author} {\bibfnamefont {A.}~\bibnamefont
  {Pimpinelli}}, \bibinfo {author} {\bibfnamefont {E.}~\bibnamefont
  {{Rastelli}}}, \ and\ \bibinfo {author} {\bibfnamefont {A.}~\bibnamefont
  {{Tassi}}},\ }\href@noop {} {\bibfield  {journal} {\bibinfo  {journal}
  {J.~Phys.:~Cond.~Mat.}\ }\textbf {\bibinfo {volume} {1}},\ \bibinfo {pages}
  {7941} (\bibinfo {year} {1989})}\BibitemShut {NoStop}%
\bibitem [{\citenamefont {Zinke}\ \emph {et~al.}(2010)\citenamefont {Zinke},
  \citenamefont {{Richter}},\ and\ \citenamefont
  {{Drechsler}}}]{Zinke2010_JPhysCondensMatter_Spiral}%
  \BibitemOpen
  \bibfield  {author} {\bibinfo {author} {\bibfnamefont {R.}~\bibnamefont
  {Zinke}}, \bibinfo {author} {\bibfnamefont {J.}~\bibnamefont {{Richter}}}, \
  and\ \bibinfo {author} {\bibfnamefont {S.-L.}\ \bibnamefont {{Drechsler}}},\
  }\href@noop {} {\bibfield  {journal} {\bibinfo  {journal}
  {J.~Phys.:~Cond.~Mat.}\ }\textbf {\bibinfo {volume} {22}},\ \bibinfo {pages}
  {446002} (\bibinfo {year} {2010})}\BibitemShut {NoStop}%
\bibitem [{elk()}]{elk}%
  \BibitemOpen
  \href@noop {} {}\bibinfo {note} {ELK FP-LAPW Code
  [http://elk.sourceforge.net/]}\BibitemShut {NoStop}%
\bibitem [{\citenamefont {Kresse}\ and\ \citenamefont
  {{Hafner}}(1993)}]{Kresse1993_PhysRevB_initio}%
  \BibitemOpen
  \bibfield  {author} {\bibinfo {author} {\bibfnamefont {G.}~\bibnamefont
  {Kresse}}\ and\ \bibinfo {author} {\bibfnamefont {J.}~\bibnamefont
  {{Hafner}}},\ }\href@noop {} {\bibfield  {journal} {\bibinfo  {journal}
  {Phys.~Rev.~B}\ }\textbf {\bibinfo {volume} {47}},\ \bibinfo {pages} {558(R)}
  (\bibinfo {year} {1993})}\BibitemShut {NoStop}%
\bibitem [{\citenamefont {Kresse}\ and\ \citenamefont
  {{Furthm}{\"u}ller}(1996)}]{Kresse1996_PhysRevB_Efficient}%
  \BibitemOpen
  \bibfield  {author} {\bibinfo {author} {\bibfnamefont {G.}~\bibnamefont
  {Kresse}}\ and\ \bibinfo {author} {\bibfnamefont {J.}~\bibnamefont
  {{Furthm}{\"u}ller}},\ }\href@noop {} {\bibfield  {journal} {\bibinfo
  {journal} {Phys.~Rev.~B}\ }\textbf {\bibinfo {volume} {54}},\ \bibinfo
  {pages} {11169} (\bibinfo {year} {1996})}\BibitemShut {NoStop}%
\bibitem [{\citenamefont {Perdew}\ and\ \citenamefont
  {{Wang}}(1992)}]{Perdew1992_PhysRevB_Accurate}%
  \BibitemOpen
  \bibfield  {author} {\bibinfo {author} {\bibfnamefont {J.~P.}\ \bibnamefont
  {Perdew}}\ and\ \bibinfo {author} {\bibfnamefont {Y.}~\bibnamefont
  {{Wang}}},\ }\href@noop {} {\bibfield  {journal} {\bibinfo  {journal}
  {Phys.~Rev.~B}\ }\textbf {\bibinfo {volume} {45}},\ \bibinfo {pages} {13244}
  (\bibinfo {year} {1992})}\BibitemShut {NoStop}%
\bibitem [{\citenamefont {Liechtenstein}\ \emph {et~al.}(1995)\citenamefont
  {Liechtenstein}, \citenamefont {{Anisimov}},\ and\ \citenamefont
  {{Zaanen}}}]{Liechtenstein1995_PhysRevB_Densityfunctional}%
  \BibitemOpen
  \bibfield  {author} {\bibinfo {author} {\bibfnamefont {A.~I.}\ \bibnamefont
  {Liechtenstein}}, \bibinfo {author} {\bibfnamefont {V.~I.}\ \bibnamefont
  {{Anisimov}}}, \ and\ \bibinfo {author} {\bibfnamefont {J.}~\bibnamefont
  {{Zaanen}}},\ }\href@noop {} {\bibfield  {journal} {\bibinfo  {journal}
  {Phys.~Rev.~B}\ }\textbf {\bibinfo {volume} {52}},\ \bibinfo {pages}
  {R5467(R)} (\bibinfo {year} {1995})}\BibitemShut {NoStop}%
\bibitem [{\citenamefont {Adachi}\ \emph {et~al.}(1972)\citenamefont {Adachi},
  \citenamefont {{Sato}}, \citenamefont {{Watarai}},\ and\ \citenamefont
  {{Ido}}}]{Adachi1972_JPhysSocJpn_Helical}%
  \BibitemOpen
  \bibfield  {author} {\bibinfo {author} {\bibfnamefont {K.}~\bibnamefont
  {Adachi}}, \bibinfo {author} {\bibfnamefont {K.}~\bibnamefont {{Sato}}},
  \bibinfo {author} {\bibfnamefont {H.}~\bibnamefont {{Watarai}}}, \ and\
  \bibinfo {author} {\bibfnamefont {T.}~\bibnamefont {{Ido}}},\ }\href@noop {}
  {\bibfield  {journal} {\bibinfo  {journal} {J.~Phys.~Soc.~Jpn.}\ }\textbf
  {\bibinfo {volume} {32}},\ \bibinfo {pages} {572} (\bibinfo {year}
  {1972})}\BibitemShut {NoStop}%
\bibitem [{\citenamefont {Hart}\ and\ \citenamefont
  {{Stanford}}(1970)}]{Hart1970_JApplPhys_Magnetic}%
  \BibitemOpen
  \bibfield  {author} {\bibinfo {author} {\bibfnamefont {L.~W.}\ \bibnamefont
  {Hart}}\ and\ \bibinfo {author} {\bibfnamefont {J.~L.}\ \bibnamefont
  {{Stanford}}},\ }\href@noop {} {\bibfield  {journal} {\bibinfo  {journal}
  {J.~Appl.~Phys.}\ }\textbf {\bibinfo {volume} {41}},\ \bibinfo {pages} {2523}
  (\bibinfo {year} {1970})}\BibitemShut {NoStop}%
\bibitem [{Note1()}]{Note1}%
  \BibitemOpen
  \bibinfo {note} {The experimental pressure studies of MnAu$_{2}$ do not find
  changes in the crystal structure prior to the spiral-to-ferromagnetic phase
  transition. For this reason, we only investigate tetragonal unit cells and do
  not consider whether other structures become competitive at higher pressures,
  since they will not impact the phase transition.}\BibitemShut {Stop}%
\bibitem [{sup()}]{supp_mater}%
  \BibitemOpen
  \href@noop {} {}\bibinfo {note} {See Supplemental Material at [URL will be
  inserted by publisher] for our raw data and a Jupyter (formerly IPython)
  notebook that describes the published research in full detail. The goal of
  including this material is so that readers can verify and reproduce all the
  presented results. The notebook contains all the Python code used to analyze
  our DFT calculations and produce the figures and tables contained in the
  manuscript. The notebook also contains some additional calculations and
  diagrams not presented in the main text that, while not essential for our
  arguments, may be of interest to the reader. Specifically, the notebook
  contains: (1) an annotated analysis of the results of structural
  optimizations performed in \textsc{vasp}, (2) a full and annotated derivation
  of the analytic results of the one-dimensional Heisenberg model, (3) details
  of how we fit our results to the one-dimensional $J_{1}-J_{2}-J_{3}-J_{4}$
  Heisenberg model, and (4) how we created the contour plots, phase diagrams,
  and visualized the exchange parameters and spiral angle $\theta$ as a
  function of pressure, doping, and $U$.}\BibitemShut {Stop}%
\bibitem [{Note2()}]{Note2}%
  \BibitemOpen
  \bibinfo {note} {Convergence of the magnetocrystalline anisotropy energy in
  the MnAu$_{2}$ unit cell was achieved with the following parameters in
  \protect \textsc {elk}: k-mesh = $28 \times 28 \times 28$, nempty = 12,
  rgkmax = 8.0, gmaxvr = 16, lmaxapw = 14, lmaxvr = 14, and the smearing
  function width was set to 0.0001 Ha}\BibitemShut {NoStop}%
\bibitem [{\citenamefont {Nagata}\ \emph {et~al.}(1999)\citenamefont {Nagata},
  \citenamefont {{Hagii}}, \citenamefont {{Samata}}, \citenamefont {{Uchida}},
  \citenamefont {{Abe}}, \citenamefont {{Fan}~{Sung}},\ and\ \citenamefont
  {{Der}~{Lan}}}]{Nagata1999_JAlComp_Magnetism}%
  \BibitemOpen
  \bibfield  {author} {\bibinfo {author} {\bibfnamefont {Y.}~\bibnamefont
  {Nagata}}, \bibinfo {author} {\bibfnamefont {T.}~\bibnamefont {{Hagii}}},
  \bibinfo {author} {\bibfnamefont {H.}~\bibnamefont {{Samata}}}, \bibinfo
  {author} {\bibfnamefont {T.}~\bibnamefont {{Uchida}}}, \bibinfo {author}
  {\bibfnamefont {S.}~\bibnamefont {{Abe}}}, \bibinfo {author} {\bibfnamefont
  {C.}~\bibnamefont {{Fan}~{Sung}}}, \ and\ \bibinfo {author} {\bibfnamefont
  {M.}~\bibnamefont {{Der}~{Lan}}},\ }\href@noop {} {\bibfield  {journal}
  {\bibinfo  {journal} {J.~Al.~Comp.}\ }\textbf {\bibinfo {volume} {284}},\
  \bibinfo {pages} {47} (\bibinfo {year} {1999})}\BibitemShut {NoStop}%
\bibitem [{\citenamefont {{Handstein}}\ \emph {et~al.}(2000)\citenamefont
  {{Handstein}}, \citenamefont {{Nenkov}}, \citenamefont {{R{\"o}ssler}},\ and\
  \citenamefont {{M{\"u}ller}}}]{Handstein2000_JAppl}%
  \BibitemOpen
  \bibfield  {author} {\bibinfo {author} {\bibfnamefont {A.}~\bibnamefont
  {{Handstein}}}, \bibinfo {author} {\bibfnamefont {K.}~\bibnamefont
  {{Nenkov}}}, \bibinfo {author} {\bibfnamefont {U.~K.}\ \bibnamefont
  {{R{\"o}ssler}}}, \ and\ \bibinfo {author} {\bibfnamefont {K.-H.}\
  \bibnamefont {{M{\"u}ller}}},\ }\href@noop {} {\bibfield  {journal} {\bibinfo
   {journal} {J.~Appl.~Phys.}\ }\textbf {\bibinfo {volume} {87}},\ \bibinfo
  {pages} {5789} (\bibinfo {year} {2000})}\BibitemShut {NoStop}%
\bibitem [{Note3()}]{Note3}%
  \BibitemOpen
  \bibinfo {note} {There is a large error in measuring $c$ as a function of
  pressure in Ref.~\protect \rev@citealp {Smith1966_JPhysChemSol_neutron}, so
  it is likely that the reported drop is not a real effect.}\BibitemShut
  {Stop}%
\bibitem [{Note4()}]{Note4}%
  \BibitemOpen
  \bibinfo {note} {We note that an alternative approach is to interpolate
  $E(\theta )$ and find the minimum, and away from phase transitions these two
  methods are in good agreement (see Supplemental Material \cite {supp_mater}).
  However, near a phase transition the energy wells become shallow and the
  interpolations are no longer reliable, so for consistency we report results
  obtained by minimizing Eq.~\protect \textup {\hbox {\mathsurround \z@
  \protect \normalfont (\ignorespaces \ref {eq:2}\unskip \@@italiccorr
  )}}}\BibitemShut {NoStop}%
\bibitem [{Note5()}]{Note5}%
  \BibitemOpen
  \bibinfo {note} {This choice of $U$ yields a $\theta < 60^{\circ }$ and a
  critical transition pressure $P \sim 20-40 \protect \text { kbar}$, in
  reasonable agreement with experiment.}\BibitemShut {Stop}%
\end{thebibliography}
\end{document}